\documentclass[%
 aip,
 amsmath,amssymb,
 reprint,%
]{revtex4-1}

\usepackage{graphicx}
\usepackage{dcolumn}
\usepackage{bm}

\usepackage[utf8]{inputenc}
\usepackage[T1]{fontenc}
\usepackage{mathptmx}
\usepackage{etoolbox}
\usepackage{color}
\makeatletter
\def\@email#1#2{%
 \endgroup
 \patchcmd{\titleblock@produce}
  {\frontmatter@RRAPformat}
  {\frontmatter@RRAPformat{\produce@RRAP{*#1\href{mailto:#2}{#2}}}\frontmatter@RRAPformat}
  {}{}
}%
\makeatother
\begin{document}
\def\be{\begin{equation}}
\def\ee{\end{equation}}

\def\bc{\begin{center}}
\def\ec{\end{center}}
\def\bea{\begin{eqnarray}}
\def\eea{\end{eqnarray}}
\newcommand{\avg}[1]{\langle{#1}\rangle}
\newcommand{\Avg}[1]{\left\langle{#1}\right\rangle}
\newcommand{\lucille}[1]{\textcolor{red}{\textbf{#1}}}
\newcommand{\lcmath}[1]{\textcolor{magenta}{\textbf{#1}}}
\newcommand{\gin}[1]{\textcolor{blue}{\textbf{#1}}}
\newcommand{\san}[1]{\textcolor{magenta}{\textbf{#1}}}
\preprint{AIP/123-QED}

\title[Local Dirac Synchronization on Networks]{Local Dirac Synchronization on  Networks}
\author{Lucille Calmon}
 \affiliation{School of Mathematical Sciences, Queen Mary University of London, London E1 4NS, UK}
 \author{Sanjukta Krishnagopal}
 \affiliation{Department of Electrical Engineering and Computer Science, University of California Berkeley, CA 94720, USA} 
\affiliation{Department of Mathematics, University of California Los Angeles, CA 90095, USA}
\author{Ginestra Bianconi}
\email{ginestra.bianconi@gmail.com}
\affiliation{School of Mathematical Sciences, Queen Mary University of London, London E1 4NS, UK}
\affiliation{The  Alan  Turing  Institute,  96  Euston  Road,  London,  NW1  2DB,  United  Kingdom}

\date{\today}

\begin{abstract}
We propose Local Dirac Synchronization which uses the Dirac operator to capture the dynamics of coupled nodes and link signals on an arbitrary network. In Local Dirac Synchronization, the harmonic modes of the dynamics oscillate freely while the other modes are interacting non-linearly, leading to a collectively synchronized state when the coupling constant of the model is increased. 
Local Dirac Synchronization  is characterized by discontinuous transitions and the emergence of a rhythmic coherent phase. In this rhythmic phase, one of the two complex order parameters oscillates in the complex plane at a slow frequency (called emergent frequency) in the  frame in which  the intrinsic frequencies have zero average. Our theoretical results obtained within the annealed approximation are validated by extensive numerical results on fully connected networks and sparse Poisson and scale-free networks.  Local Dirac Synchronization on both random and real networks, such as the connectome of Caenorhabditis Elegans, reveals the interplay between topology (Betti numbers and harmonic modes) and non-linear dynamics. This unveils how topology  might play a role in the onset of brain rhythms. 
\end{abstract}
\maketitle

\begin{quotation}
Topological signals are dynamical variables associated to nodes and  links of a network, as well as to higher dimensional simplices of simplicial complexes. Topological signals are attracting increasing attention in brain research and data science, however the investigation of their collective dynamics is only at its infancy. Here we use the Dirac operator, an operator inspired by quantum mechanics but fully grounded in algebraic topology, to propose Local Dirac Synchronization involving coupled nodes and link signals on an arbitrary network. These results open new perspectives on the interplay between structure and dynamics of networks and demonstrate the key role of topology (Betti numbers and harmonic modes) in determining the dynamical properties of topological signals on arbitrary network structures. This work focuses exclusively on the dynamics of coupled topological signals defined on  networks (nodes and link signals), however Local Dirac Synchronization can be extended to treat coupled dynamics of higher-order topological signals on simplicial complexes.
\end{quotation}

\section{Introduction}
Networks  \cite{barabasi2016network,newman2018networks,dorogovtsev2022nature} are key to uncover the underlying topology of {a variety of} interacting complex systems ranging from the brain to power-grids. However, in order to predict the function of complex systems from their network structure, it is of fundamental importance to understand the interplay between their structure and their dynamics.  
In the last two decades there has been great progress in revealing how the combinatorial and  statistical properties  of networks (such as their degree distribution) affect critical phenomena defined on them \cite{dorogovtsev2008critical}. Notably, it has been shown that  scale-free degree distributions change the phase diagram of percolation, the Ising model and epidemic spreading with notable effects on the collective dynamical behavior of networks.
Interestingly, synchronization processes taking place on  networks such as the Kuramoto model \cite{strogatz2000kuramoto,arenas2008synchronization,rodrigues2016kuramoto,gross2021not} are affected non-trivially by heterogeneous degree distributions and network symmetries \cite{pecora2014cluster,morone2020fibration,salova2021cluster,hart2019topological} as well.
The spectral properties of networks are also known to determine their synchronization dynamics. In the context of global synchronization of identical oscillators, most works have so far focused, almost exclusively, on the role of the Fiedler eigenvalue of the graph Laplacian in determining the stability of global synchronization \cite{pecora1998master,barahona2002synchronization,mulas2020coupled}. {  However for Kuramoto-like models more general spectral properties have also been shown to play a key role \cite{skardal2014optimal,tyloo2018robustness,ronellenfitsch2018optimal}. }

Recently, sparked by the growing interest on higher-order networks \cite{battiston2020networks,bianconi2021higher,bick2021higher},  the investigation of the interplay between the topology of simple or higher-order networks and their associated dynamics \cite{bianconi2021higher,millan2020explosive,ghorbanchian2020higher,torres2020simplicial,battiston2021physics,majhi2022dynamics} is gaining growing attention.
Topology is the study of shapes and their invariant properties that do not depend on metric considerations. A notable set of topological invariants are the Betti numbers that enumerate the number of cavities in  the simple or  higher-order network under consideration.
For instance, in a network, the  Betti number $\beta_0$ indicates the number of connected components while the  Betti number  $\beta_1$ indicates the number of independent cycles in the network.
Interestingly, topology can be used not only to study the structure of networks \cite{giusti2016two,otter2017roadmap,patania2017topological,krishnagopal2021spectral} but  it can also be used also to investigate their higher-order dynamics \cite{millan2020explosive,bianconi2021topological,torres2020simplicial,ghorbanchian2020higher,millan2021geometry}.
Indeed, a network dynamical state in general will not only include dynamical variables associated to its nodes, also called node signals (already widely studied in the literature) but will also include  link signals, i.e. dynamical variables associated to the links.
Node and link signals of a network are examples of topological signals and are attracting large attention in signal processing and data science \cite{Barbarossa,schaub2020random,schaub2021signal}.
Topological signals might include the synaptic signals between neurons or the edge signals between macroscopic regions of the brain \cite{faskowitz2022edges}, as well as biological transportation fluxes \cite{rocks2021hidden}, currents in power-grids \cite{kaiser2021topological}, or even speed of wind or currents at a given location on the surface of the Earth or of the ocean \cite{schaub2020random}.
On simplicial and cell complexes, topological signals can {also} be defined  on higher dimensional simplices and cells such as triangles, squares, tetrahedra, cubes  and so on.

Recently, it was shown \cite{millan2020explosive,ghorbanchian2020higher,millan2021geometry} that topological signals of a given dimension (for instance defined on links or on triangles), can undergo collective phenomena. In particular, they can undergo (partial) synchronization transitions in the framework of the topological higher-order Kuramoto model \cite{millan2020explosive}, where the collective dynamics is localized into the harmonic modes of the dynamics. 
Interestingly, the topological higher-order dynamics has  been extended to treat the case in which the higher-order networks are weighted or directed \cite{deville2020consensus,arnaudon2022connecting}.

In real complex systems, both partial and global synchronization can occur in different scenarios. Partial synchronization  for instance captures brain dynamics while global synchronization is known to occur in other complex systems such as power-grids. Recently it was shown \cite{carletti2022global} that  global synchronization of topological signals can also be achieved under suitable topological and spectral conditions. These conditions include the notable constraint that the harmonic eigenvector of the higher-order Laplacian is has elements of constant absolute value, which is satisfied, for instance, on square lattices of arbitrary dimension.

However, most works so far focus exclusively on the collective dynamics of topological signals of a single dimension, for instance, on synchronization of link signals. Since in a network, node signals and link signals might coexist, an important question is whether their dynamics can be coupled in non-trivial ways. To couple node and link dynamics, the topological Dirac operator \cite{bianconi2021topological,bianconi2022dirac,baccini2022weighted,lloyd2016quantum,calmon2023dirac} is particularly well-suited. Indeed, the topological Dirac operator allows for cross-talking between nodes and link signals by projecting one onto the other, allowing the treatment of coupled dynamics of topological signals of different dimensions \cite{calmon2021topological,giambagli2022diffusion}.
In Ref. \cite{calmon2021topological} the Dirac operator was used for the first time to formulate Dirac synchronization on a fully connected network. Dirac synchronization uses the Dirac operator to couple the dynamics of oscillators placed on nodes and links of a network. The resulting dynamics displays a collective phenomenon in which nodes and link signals are intertwined and the coherent phase is rhythmic. These results might therefore reveal possible topological mechanisms for the emergence of brain rhythms \cite{breakspear2010generative,buzsaki2006rhythms}.

\begin{figure}[tbh]
\centering
\includegraphics[width=1\columnwidth]{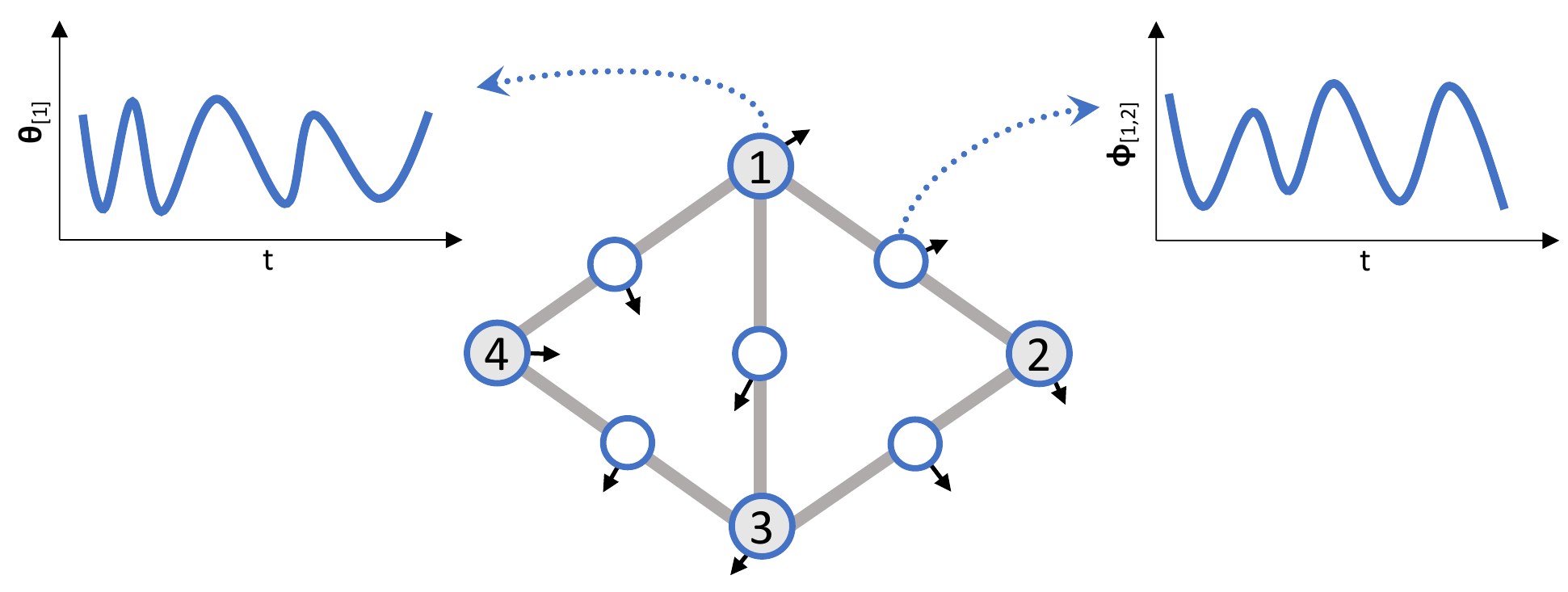}
\caption{The Dirac operator couples locally and topologically signals  defined on nodes and links. Local Dirac Synchronization on an arbitrary network  couples oscillators located on nodes and links through a local phase lag, thus inducing a highly non-trivial feedback loop between the two types of variables.
}
\label{fig:1}       
\end{figure}

In this work we investigate the normalized Dirac operator \cite{baccini2022weighted} which generalizes the  un-normalized Dirac operator \cite{bianconi2021topological}, and show that this allows for {\em Local Dirac Synchronization} (LDS), which treats the coupled dynamics of node and link signals on a connected but otherwise arbitrary network. The model is called {\em local} because the normalized Dirac operator couples the dynamics of nodes  to the dynamics of nearby links though, and vice versa, couples the dynamics of links to nearby nodes {  where the nodes and links are coupled through  phase lags} (see Fig. $\ref{fig:1}$). 
Moreover, LDS is controlled by the parameter $z$ which can tune the strength of phase-lags, changing the phase diagram of the model. Indeed, {  we observe that} by varying $z$, the backward transition of the model  changes from continuous to discontinuous. These behaviors provide new perspectives into the very vibrant debate about discontinuous synchronization transitions in simple, multiplex and higher-order networks \cite{Ott_Girvan,skardal2019abrupt,skardal2020higher,boccaletti2016explosiveReview,d2019explosive,zou2014basin,dai2020discontinuous,sarika_2022hebbian,sarika,lucas2020multiorder}, and demonstrate how topology can be a key player for the onset of abrupt transitions. Indeed, in this scenario, the local coupling induced by the normalized Dirac operator in the form of phase-lags induces a phase transition reminiscent of the one of the  Sakaguchi-Kuramoto model \cite{sakaguchi1986soluble,omel2012nonuniversal,omel2013bifurcations}. Note however that while in the Sakaguchi-Kuramoto model the phase-lag is constant, in LDS, it is  adaptive.
Lastly, we present a theoretical analysis, based on the annealed approximation, to predict the phase diagram of LDS on uncorrelated networks, {  which we validate with} extensive numerical simulations.
It is important to note that the coherent phase of LDS is a  rhythmic phase and can be observed not only on fully connected networks but also on sparse uncorrelated random networks such as Poisson and scale-free networks.

This paper is structured as follows: in Sec. II we define topological signals associated with nodes and links and outline the main spectral properties of the normalized and un-normalized Dirac operators; in Sec. III we provide a review of the main results of the Topological Kuramoto model which treats the synchronization of uncoupled nodes and link signals; in Sec. IV we introduce the LDS model; In Sec. V  we discuss the main properties of the LDS model including its non-trivial phase diagram and the emergence of the rhythmic phase;
in Sec. VI we investigate in details the phase diagram of LDS on fully connected as well as on sparse Poisson and scale-free networks comparing our theoretical results to extensive numerical simulations. {  Moreover we study LDS on the real connectome of C.elegans finding evidence for a discontinuous phase transition and a rhythmic phase.} Finally in Sec. VII we provide the concluding remarks. The paper  includes three Appendices providing the details of our analytical derivations.

\section{Topological signals and Dirac operators}
Consider a network $G=(V,E)$ comprising of a set $V$ of $N$ nodes and  a set $E$ of $L$ links. The links have an orientation induced by the node labels, i.e. the link $[i,j]$ has positive orientation if and only if $i<j$. The network's topology is fully characterised by the $N\times L$ boundary matrix ${\bf B}$, which provides a map from the positively oriented links to their respective endpoints. Specifically the matrix elements of ${\bf B}$ are given by
\begin{equation}
{B}_{i\ell}=\left\{\begin{array}{cc}1 & \mbox{if}\   \ell=[j,i], \ j<i\\ -1 & \mbox{if} \ \ell=[i,j],\  i<j\\  0 &   \mbox{otherwise}.\end{array}\right.
\label{B_link}
\end{equation}
where here $\ell$ indicates a generic positively oriented link.
The Dirac operator of simplicial complexes, introduced in Ref.~\cite{bianconi2021topological}, provides the natural framework to couple topological signals of different orders together. On a network, the Dirac operator is  represented by the $(N+L)\times (N+L)$ matrix ${\bf D}$ with the following block structure: 
\begin{equation}
\bf D=\left(\begin{array}{cc}{\bf 0}& {\bf B}\\
{\bf B}^{\top}&{\bf 0}\end{array}\right).
\end{equation}
The Dirac operator has two fundamental properties. First, it can be interpreted as the "square root" of the super-Laplacian $\mathcal{L}$, indeed the square of the Dirac operator 
is given by
\bea
{\bf D}^2=\mathcal{{{L}}}=\begin{pmatrix}{\bf L}_{0} &{\bf 0}\\
{\bf 0}&{\bf L}_{1} \end{pmatrix},
\eea
where ${\bf L}_{0}={\bf B}{\bf B}^{\top}$, ${\bf L}_{1}={\bf B}^{\top}{\bf B}$ are respectively the graph Laplacian and the first-order Laplacian of the considered  network.
The second fundamental property of the Dirac operator is that it  allows cross-talking between signals defined on nodes and links.
Indeed the Dirac operator acts on  topological spinors of the type \bea
\bm\Phi=\begin{pmatrix}\bm \theta\\\bm \phi\end{pmatrix}
\eea
where $\bm\theta$ indicates a 0-cochain defined on the nodes of the network, and $\bm\phi$ indicates a $1$-cochain defined on the links of the network.
For instance $\bm \theta$ can indicate the phases of the oscillators placed on the nodes and $\bm \phi$ can indicate the phases of the oscillators placed on the links of the network.
When we apply the Dirac operator to this topological spinor, we obtain 
\bea
{\bf D}\bm \Phi=\begin{pmatrix}{\bf B}\bm \phi\\{\bf B}^{\top}\bm \theta\end{pmatrix}
\eea
where ${\bf B}\bm\phi$ indicates the projection of the links signal onto the nodes (divergence of the links signal) and ${\bf B}^{\top} \bm\theta$ indicates the projection of the nodes signal onto the links (gradient).
This second property implies that the Dirac operator can be used to study the coupled dynamics defined on simplices of different dimension, which, in a network, translate into the coupled dynamics of signals associated with nodes and links.

Since the square of the Dirac operator is the super-Laplacian matrix $\mathcal{L}$, and the spectrum of $\mathcal{L}$ is given by the concatenation of the spectrum of ${\bf L}_0$ and ${\bf L}_1$ (which are isospectral), it can be easily shown that the eigenvalues $\lambda$ of the Dirac operator are given by the square roots of the eigenvalues of the graph Laplacian ${\bf L}_{0}$ taken with both positive and negative sign, i.e.
\bea
\lambda=\pm\sqrt{\mu},
\label{Ds}
\eea
where $\mu$ indicates the generic eigenvalue of the graph Laplacian ${\bf L}_0$.
The matrix $\bm\Psi$  of eigenvectors of the  Dirac operator ${\bf D}$ is given by  
\bea
\bm\Psi=\begin{pmatrix}{\bf {U}}&{\bf U}&{\bf {U}}_{harm},{\bf 0}\\
{\bf {V}}&-{\bf {V}}&{\bf 0}&{\bf V}_{harm}
\end{pmatrix},
\eea
where ${\bf U}$ and ${\bf V}$ are the matrices of the left and right singular vectors corresponding to non-zero singular values of ${\bf B}$ respectively while ${\bf U}_{harm}$ and ${\bf V}_{harm}$ are in the kernel of ${\bf B}^{\top}$ and ${\bf B}$ respectively.
Here the  matrices of eigenvectors  $({\bf {U}},{\bf {V}})^{\top}$ and $({\bf {U}},-{\bf {V}})^{\top}$ correspond to the positive and negative eigenvalues of the  Dirac operator ${\bf D}$  respectively. For each pair of eigenvalues $\lambda,-\lambda$ with $\lambda\neq 0 $, these eigenvectors are  therefore chiral to each other. The matrices of  harmonic eigenvectors associated with the eigenvalue $\lambda=0$ are instead  $({\bf U}_{harm},{\bf 0})^{\top}$ and $({\bf 0},{\bf V}_{harm})^{\top}$.
The degeneracy of the zero eigenvalues of the Dirac operator ${\bf D}$ is given by the sum of Betti numbers $\beta_0+\beta_1$ with the harmonic eigenvectors of the type $({\bf U}_{harm},{\bf 0})^{\top}$ having a $\beta_0$ degeneracy and the harmonic eigenvectors of the type $({\bf 0},{\bf V}_{harm})^{\top}$ having degeneracy $\beta_1$.

Since the spectrum of the Dirac operator is related to the spectrum of the graph Laplacian by Eq. (\ref{Ds}), as long as the maximum degree of the network is not bounded, the eigenvalues of the Dirac operators are not bounded either.
However, in a number of cases, such as for the present study of Dirac synchronization, it is not only convenient, but in fact  necessary to consider a normalized Dirac operator \cite{baccini2022weighted} with bounded eigenvalues.

Here we consider the normalized Dirac operator $\hat{\bf D}$ given by 
\begin{equation}
\hat{\bf D}={\mathcal{K}}^{-1}\bf{D}
\end{equation}
where $\mathcal{K}$ is the matrix of (generalized) degrees of nodes and links given by 
\begin{equation}
\boldsymbol{\mathcal{K}}=\left(\begin{array}{cc}{\bf K}_0& {\bf 0}\\
{\bf 0}&{\bf K}_1\end{array}\right),
\end{equation}
where ${\bf K}_0$ is the $N \times N$ diagonal matrix with diagonal elements given by the node degrees ($K_{0}(i,i)=k_i$) and ${\bf K}_1$ is the $L \times L$ diagonal matrix whose diagonal elements are given by  the  down degrees of the links ($K_{1}(i,i)=2$).
It follows that the normalized Dirac operator can be explicitly expressed as
\begin{equation}
\hat{\bf D}=\begin{pmatrix}{\bf 0}&{\bf K}_0^{-1}{\bf B}\\{\bf B}^{\top}/2&{\bf 0}\end{pmatrix}.
\end{equation}
Interestingly the square of the normalized Dirac operator is given by the normalized super-Laplacian
\bea
\hat{\bf D}^2=\hat{\boldsymbol{\mathcal{L}}}=\left(\begin{array}{cc}\hat{\bf L}_{0}/2 &{\bf 0}\\
{\bf 0}&\hat{\bf L}_{1}/2 \end{array}\right),
\eea
where $\hat{\bf L}_{0}={\bf K}_0^{-1}{\bf B}{\bf B}^{\top}$ is the normalized graph Laplacian and 
$\hat{\bf L}_{1}={\bf B}^{\top}{\bf K}_0^{-1}{\bf B}$ the normalized $1$-st order Laplacian on the considered network.

A well known result of spectral graph theory \cite{chung} is that  $\hat{\bf L}_{0}$ and $\hat{\bf L}_{1}$ are isospectral and have eigenvalues smaller or equal to $2$.
It follows that both $\hat{\bf L}_{0}/2$ and $\hat{\bf L}_{1}/2$
have spectrum bounded from above by one, and consequently the normalized Dirac operator $\hat{\bf D}$ also has eigenvalues whose absolute value is bounded by one and given by 
\bea
\lambda=\pm \sqrt{\mu}
\eea
where $\mu$ indicates the generic eigenvalue  of $\hat{\bf L}_0/2$ and $|\lambda|\leq 1$.
The normalized Dirac operator being no longer symmetric, we  distinguish between the  right and left eigenvectors.
The matrix $\bm{\hat{\Psi}}^{L}$ of left  eigenvectors of the normalized Dirac operator is given by  
\bea
\bm{\hat{\Psi}}^{L}=\begin{pmatrix}{\bf \hat{U}}^L&{\bf \hat{U}}^L&{\bf \hat{U}}_{harm}^L,{\bf 0}\\
{\bf \hat{V}}^L&-{\bf \hat{V}}^L&{\bf 0}&{\bf \hat{V}}_{harm}^L
\end{pmatrix}.
\eea
The left eigenvectors  $({\bf \hat{U}}^L,{\bf \hat{V}}^L)^{\top}$ and $({\bf \hat{U}}^L,-{\bf \hat{V}}^L)^{\top}$ correspond, respectively, to the positive and negative eigenvalues $\pm|\lambda|$ of the normalized Dirac operator $\hat{\bf D}$. For each pair of eigenvalues $\lambda, -\lambda$, the corresponding eigenvectors are related by chirality. The left harmonic eigenvectors $({\bf \hat{U}}_{harm}^L,{\bf 0})^{\top}$ and $({\bf 0},{\bf \hat{V}}_{harm}^L)^{\top}$ where ${\bf \hat{U}}_{harm}^L$  {is} the matrix of  left harmonic singular vectors of ${\bf K}_0^{-1}{\bf B}$ and ${\bf \hat{V}}_{harm}^L$ is the matrix of left harmonic singular vectors of ${\bf B}^{\top}/2$, are associated with the eigenvalue $\lambda=0$.
Here ${\bf \hat{U}}^L$ indicates the set of vectors ${\bf \hat{u}}^L_{\lambda}={\mathcal{C}}{\bf K}_{[0]}^{-1/2}\tilde{\bf u}_{\lambda}$ and ${\bf \hat{V}}$ indicates the set of vectors ${\bf \hat{v}}_{\lambda}^L={\mathcal{C}}\tilde{\bf v}_{\lambda}/\sqrt{2}$ where $\tilde{\bf u}_{\lambda}$ and $\tilde{\bf v}_{\lambda}$ are respectively the right and left singular vectors of  ${\bf K}_0^{-1/2}{\bf B}_{[1]}/\sqrt{2}$  corresponding to the singular value $|\lambda|\neq 0$ and $\mathcal{C}$ is a normalization constant.
Note that while the eigenvectors associated with the non-zero eigenvalues of the normalized Dirac operators have non-zero elements on both nodes and links, the harmonic eigenvectors can be chosen in such a way that they are either localized on nodes or on links.

The matrix $\bm{\hat{\Psi}}^{R}$ of right  eigenvectors of the normalized Dirac operator has a similar structure and can be expressed as 
\bea
\bm{\hat{\Psi}}^{R}=\begin{pmatrix}{\bf \hat{U}}^R&{\bf \hat{U}}^R&{\bf \hat{U}}_{harm}^{R}&{\bf 0}\\
{\bf \hat{V}}^R&-{\bf \hat{V}}^R&{\bf 0}&{\bf \hat{V}}^{R}_{harm}
\end{pmatrix},
\eea
where ${{\bf \hat{U}}_{harm}}^R$ is the matrix of right harmonic singular vectors of ${\bf K}_0^{-1}{\bf B}$ and ${\bf \hat{V}}_{harm}^L$ is the matrix of right harmonic singular vectors of ${\bf B}^{\top}/2$.
Moreover ${\bf \hat{U}}^R$ indicates the set of vectors ${\bf \hat{u}}^R_{\lambda}={\mathcal{C}}{\bf K}_{[0]}^{1/2}\tilde{\bf u}_{\lambda}$ and ${\bf \hat{V}}^R$ indicates the set of vectors ${\bf \hat{v}}_{\lambda}^R={\mathcal{C}}\tilde{\bf v}_{\lambda}\sqrt{2}$ where $\tilde{\bf u}_{\lambda}$ and $\tilde{\bf v}_{\lambda}$ are respectively the right and left singular vectors of ${\bf K}_0^{-1/2}{\bf B}_{[1]}/\sqrt{2}$ corresponding to its singular value $|\lambda|\neq 0$ and $\mathcal{C}$ is a normalization constant.

\section{Topological Kuramoto model}
The Kuramoto model is a paradigmatic model to study synchronization of oscillators placed on the nodes of a network.
These oscillators are non-identical, and in the absence of interactions, each oscillator $i$ oscillates at a frequency $\omega_i$ typically drawn from a unimodal distribution such as a Gaussian or a Lorentzian. The Kuramoto model then introduces a coupling between oscillators placed on nearby nodes which is modulated by the coupling constant $\sigma>0$.
Denoting the phases of these oscillators placed on the nodes of the network by the vector $\bm \theta=(\theta_1,\theta_2,...\theta_N)^{\top}$, and indicating their intrinsic frequencies by the vector $\bm\omega=(\omega_1,\omega_2,...\omega_N)^{\top}$, the Kuramoto model \cite{kuramoto1975,strogatz2000kuramoto,rodrigues2016kuramoto} is a dynamical system dictated by the following system of equations, 
\begin{equation}
    \dot{\bm{\theta}} = \bm{\omega} - \sigma{\bf B}\sin({
   \bf B}^\top\bm{\theta}),
   \label{Kuramoto_std}
\end{equation}
where here and in the following $\sin(\bf{x})$ denotes the vector where the $\sin$ function is taken element wise. 
Interestingly this model displays a synchronization phase transition. For low values of $\sigma$ there is no collective dynamics while for $\sigma $ above the synchronization threshold $\sigma_c$, i.e. $\sigma>\sigma_c$, a  synchronized state emerges in which a finite fraction of the nodes oscillates at the same frequency.
Note that from Eq.(\ref{Kuramoto_std}) it is apparent that if we project the dynamics of the phases $\bm\theta$ on the eigenvector on the kernel of ${\bf B}^{\top}$, i.e. if we project $\bm\theta$ into the  harmonic eigenvector ${\bf u}_{harm}=\mathcal{C}{\bf 1}$ (with  ${\bf 1}$ being the vector taking value $1$ on each node of the network and $\mathcal{C}$ being a normalization constant), we obtain
\bea
\frac{d\langle {\bf u}_{harm}|\bm \theta \rangle}{dt} =\langle {\bf u}_{harm}|\bm \omega\rangle.
\eea
This equation implies that the projected dynamics onto the harmonic mode(s) oscillates independently of the remaining non-harmonic components of $\bm\theta$. Furthermore its own intrinsic frequency remains unaffected by the coupling constant.
Therefore the synchronization transition can be understood as the  freezing of the non-harmonic modes of the dynamics, while the mode determined by $\langle {\bf u}_{harm}|\bm \theta \rangle$ plays a dominant role on the resulting collective dynamics of the system.
The macroscopic state of this system is captured by the global order parameter
\bea
    R_{\theta} = \frac{1}{N}\left|\sum_{j=1}^{N}e^{\textrm{i}\theta_j}\right|
\eea
which takes values from zero to one. The synchronization threshold $\sigma_c$ indicates the value of the coupling constant at which the order parameter $R_{\theta}$ becomes non-zero  in the limit $N\to\infty$.
    
Recently, great interest has been devoted to the topological Kuramoto model \cite{millan2020explosive} in which the oscillators can be placed on links as well as higher dimensional simplices of a simplicial complex.
On a network, this topological higher-order Kuramoto model describes the synchronization of oscillators placed on the links of the network and coupled together through shared nodes.
The topological higher-order Kuramoto model is dictated by the system of equations
\begin{equation}
    \dot{\bm{\phi}} = \bm{\tilde{\omega}} - \sigma{\bf B}^{\top}\sin({
   \bf B}\bm{\phi})
   \label{Kuramoto_higher}
\end{equation}
which captures the synchronization of the phases $\bm\phi=(\phi_1,\phi_2,\ldots, \phi_{\ell},\ldots \phi_L)^{\top}$ of the oscillators placed on the links with intrinsic frequencies $\bm{\tilde{\omega}}=({\tilde{\omega}}_1,{\tilde{\omega}}_2,\ldots,{\tilde{\omega}}_{\ell},\ldots, {\tilde{\omega}}_L)$. 

In this case, each projection of the phases $\bm\phi$ onto the harmonic eigenvectors ${\bf v}_{harm}$ of ${\bf B}^{\top}{\bf B}$ 
will dominate the dynamics while the other modes will freeze progressively. Indeed we have that for each harmonic eigenvector ${\bf v}_{harm}$, the dynamics of the mode $\langle {\bf v}_{harm}|\bm \phi \rangle$ obeys
\bea
\frac{d\langle {\bf v}_{harm}|\bm \phi \rangle}{dt} =\langle {\bf v}_{harm}|\bm \hat{\omega}\rangle
\eea
while the progressing freezing of the other modes 
leads to the higher-order topological synchronization of links oscillators which was shown in Ref. \cite{millan2020explosive}  to display a synchronization threshold $\sigma_c=0$.
The freezing of the non-harmonic modes of the dynamics can be monitored by projecting the phases of the links $\bm\phi$ onto the nodes with the transformation
\bea
\bm\psi={\bf B}\bm\phi,
\eea
which screens out the contributions of the harmonic modes.
The dynamical equation for $\bm \psi$ can be obtained directly from Eq. (\ref{Kuramoto_higher}) by multiplying both sides by ${\bf B}$, yielding
\bea
\dot{\bm\psi}=\bm\hat{\omega}-\sigma {\bf L}_0\sin(\bm\psi).
\eea
where $\bm\hat{\omega}={\bf B}\bm\tilde{\omega}$.
Hence, the synchronization of the oscillators of the links can be monitored by the order parameter 
\bea
    R_{\psi} = \frac{1}{N}\left|\sum_{j=1}^{N}e^{\textrm{i}\psi_j}\right|.
\eea

The independent synchronization of oscillators associated with the nodes (Eq. (\ref{Kuramoto_std})) and to the links (Eq. (\ref{Kuramoto_higher})) of the same network can therefore be expressed in terms of the topological spinor $\bm \Phi=(\bm\theta,\bm\phi)^{\top}$ and the Dirac operator ${\bf D}$ as 
\begin{equation}
    \dot{\bm{\Phi}} = \bm{\Omega} - \sigma{\bf D} \sin({
   \bf D}\bm{\Phi}),
   \label{Kuramoto_uncoupled}
\end{equation}
where $\Omega=(\bm\omega,\bm\tilde{\omega})^{\top}$ indicates the topological spinor of intrinsic frequencies of the nodes ($\bm\omega$) and of the links ($\bm{\tilde\omega}$). 
This equation is totally equivalent to Eqs. (\ref{Kuramoto_std}) and (\ref{Kuramoto_higher}). Interestingly however this system of equations can be generalized by replacing the Dirac operator ${\bf D}$ with its normalized counterpart ${\bf \hat{D}}$, leading to
\begin{equation}
    \dot{\bm{\Phi}} = \bm{\Omega} - \sigma \hat{\bf D} \sin(\hat{
   \bf D}\bm{\Phi}).
   \label{Kuramoto_uncoupled_2}
\end{equation}
This dynamics is equivalent to the following uncoupled dynamics for nodes and link oscillators
\bea
    \dot{\bm{\theta}} &=& \bm{\omega} - \sigma{\bf K}_0^{-1}{\bf B}\sin({
   \bf B}^\top\bm{\theta}/2),\\
\dot{\bm{\phi}} &=& \bm{\tilde{\omega}} - \frac{1}{2}\sigma{\bf B}^{\top}\sin({\bf K}_0^{-1}{
   \bf B}\bm{\phi}).
\eea
This  dynamics  describes a different physical problem than Eq. (\ref{Kuramoto_uncoupled}) as it screens out the eventual effect of  heterogeneous degree distributions. Indeed the resulting dynamics for the nodes oscillators $\bm\theta$ is a very popular choice when one desires to reduce the effect of broad degree distributions, or to reveal the intrinsic effects of discrete geometry on synchronization dynamics \cite{millan2019synchronization,millan2021geometry}.
Both dynamics dictated by Eq. (\ref{Kuramoto_uncoupled}) and Eq. $(\ref{Kuramoto_uncoupled_2})$ describe the uncoupled dynamics of oscillators placed on the nodes and on the links of the considered network.

The investigation of the linearization of these two dynamics can provide useful insight into the topological Kuramoto model.
The linearized dynamics of Eq. (\ref{Kuramoto_uncoupled}) driven by the super-Laplacian $\mathcal{L}$ is given by 
\bea
 \dot{\bm{\Phi}} = \bm{\Omega} - \sigma {\boldsymbol{\mathcal{L}}} \bm{\Phi}
 \label{TKlin_unnorm}
\eea
while the linearized dynamics of Eq.(\ref{Kuramoto_uncoupled_2}) is driven by the normalized super-Laplacian $\mathcal{\hat{L}}$ and is given by 
\bea
 \dot{\bm{\Phi}} = \bm{\Omega} - \sigma \mathcal{\hat{L}} \bm{\Phi}
 \label{TKlin_norm}
\eea
Interestingly in these two linearized dynamics (see Appendix \ref{AppendixA} for details), the harmonic modes are decoupled, and their dynamics only depend on the intrinsic frequencies, while the non-harmonic modes asymptotically freeze, i.e. they converge to a constant. Moreover, we note that as expected, the dynamics of the node oscillators is decoupled from the dynamics of the links oscillators. This is due to the symmetry of the (normalized and un-normalized) Dirac eigenvalues and eigenvectors.

The topological higher-order Kuramoto model that we have studied in this section is able to capture the collective synchronization phenomena of non-identical oscillators placed on the nodes and links of the same network.
However, so far, these two dynamics remain uncoupled. An important question of that emerges in this context is whether the dynamics of nodes and links can be dynamically coupled.

\section{Local Dirac synchronization}
\label{Local  Dirac synchronization}
\subsection{General formulation of Dirac synchronization on an arbitrary network}

\emph{Local Dirac synchronization} (LDS) is a   model that allows the coupling of nodes and link signals locally and topologically on an arbitrary connected network.
The dynamics of the phases of the nodes are  coupled to the phases of neighbouring links via a phase lag.  {Vice versa}, the dynamics of the oscillators on links is affected by a phase lag determined by the phases of neighbouring adjacent nodes. This phase-lag, reminiscent of the uniform added constant in the Sakaguchi-Kuramoto model~\cite{sakaguchi1986soluble}, is modulated by the parameter $z$. {  In contrast with} the Sakaguchi-Kuramoto model which employs a constant uniform 
phase-lag, here, our topological coupling through the (normalized) Dirac operator instead implements a non-trivial feedback mechanism between the two types of oscillators. {  It is important to note that }this dynamical phase-lag depends on the neighbouring adjacent phases, and therefore is time dependent. 

Dirac synchronization on a  network of arbitrary topology is defined as a function of a fixed parameter $z$ and obeys the dynamical equations
\begin{equation}
    \dot{\bm{\Phi}} = \bm{\Omega}-\sigma \hat{\bf D}\sin \left(\hat{\bf D}\bm{\Phi}-z{\bm \gamma}\hat{\bf D}^2\bm{\Phi}\right)
    \label{Dirac_synchr}
\end{equation}
where $\bm\Phi=(\bm\theta,\bm\phi)^{\top}$ is the topological spinor indicating the phases of the nodes and of the links of the network and $\bm\Omega=(\bm\omega,\bm\tilde{\omega})^{\top}$ is the vector of intrinsic frequencies of  nodes and  links.
In Eq. (\ref{Dirac_synchr}), the matrix $ \bm\gamma$ takes the form
\begin{equation}
 \bm{\gamma}=\left(\begin{array}{cc}{{\bf I}_{N}}& {\bf 0}\\
{\bf 0}&{{-\bf I}_{L}}\end{array}\right)
\label{gamma}
\end{equation}
where ${\bf I}_Y$ is the $Y \times Y$ identity matrix.
Note that due to the chirality of the spectrum of the normalized Dirac operator $\hat{\bf D}$, there is a symmetry between the dynamics observed for positive and negative values of $z$. Therefore, without loss of generality, we can restrict our study to the dynamics obtained for $z>0$.

Let us now discuss and motivate the implications of the dynamics of Dirac synchronization.
First of all we notice that the harmonic modes of the dynamics oscillate freely with a frequency dictated by the intrinsic frequencies.
Indeed if we denote by $\bm\Phi_{harm}$ the projection of $\bm\Phi$ onto a generic right harmonic eigenvector of the normalized Dirac operator, and by $\bm\Omega_{harm}$ the projection of $\bm\Omega$  {o}nto the same right harmonic eigenvector, we can easily show that Eq. (\ref{Dirac_synchr}) implies 
\bea
\dot{\bm\Phi}_{harm}=\bm\Omega_{harm}.
\eea
However the other modes of the dynamics are coupled together by the interaction term.
The choice of the interaction term can be discussed and motivated by considering the linearized dynamic of the model.
The  linearized dynamics of Eq. (\ref{Dirac_synchr}) close to the fully synchronized state takes the form (note that $\bm\gamma$ and $\bf \hat{D}$ anticommute)
\begin{equation}
    \dot{\bm{\Phi}} = \bm{\Omega}-\sigma (\hat{\bf D}^2+z{\bm \gamma}\hat{\bf D}^3)\bm\Phi.
    \label{Dirac_linearised}
\end{equation}
The adopted expression for $\bm\gamma$ allows us to couple the chiral eigenvectors corresponding to positive and negative eigenvalues $\pm\lambda\neq 0$ with the same absolute value.
From the study of the linearized dynamics (Eq.(\ref{Dirac_linearised})), we can make two important observations on the opportune choice of the matrix $\bm\gamma$ (see Appendix\ref{AppendixA} for details).
First of all with the adopted expression of the matrix $\bm\gamma$, the linearized operator $\hat{\bf D}^2+z{\bm \gamma}\hat{\bf D}^3$ admits for any choice of $z>0$ eigenvalues  {with} positive real part, therefore describing the damping of non-harmonic modes.
Note however that if we were to substitute the identity matrix  in place of $\bm\gamma$, $\hat{\bf D}^2+z{\bm \gamma}\hat{\bf D}^3$  would admit negative eigenvalues for $|z|>1$.
Secondly, this choice of the matrix $\bm\gamma$ leads to (emergent) damped oscillations of the non-harmonic modes of the dynamics, which are absent if we replace $\bm\gamma$ by the identity matrix. Therefore we intuitively believe that these emergent oscillations can be related to the stable (emergent) oscillations which constitute a characteristic feature of Dirac synchronization that we  discuss in the next sections.

On a fully connected network Dirac synchronization for $z=1$ (up to a factor of $2$) has been investigated in detail in  Ref.~\cite{calmon2021topological}. There it was shown that the phase diagram of Dirac synchronization displays a discontinuous forward transition and a continuous backward transition, as well as a rhythmic phase in which one order parameter of the dynamics oscillates at an emergent frequency in the rotating frame of the intrinsic frequencies.
However in Ref.~\cite{calmon2021topological} the formulation of the model was in terms of the un-normalized Dirac.
In this paper, the new formulation of the Dirac synchronization dynamics in terms of the normalized Dirac operator allows us to formulate LDS and to directly use this model to study the coupled dynamics of nodes and link signals on any general network topology.  Furthermore, the parameter $z$ present in LDS and not initially present in Dirac synchronization model formulated in  Ref.~\cite{calmon2021topological}, allows a further tuning of the coupling term. As we show in the remainder, the phenomenology of both models on fully connected networks are similar for small enough $z$. However, increasing $z$ leads to a discontinuous backward transition as well, which is a feature reminiscent of the phase diagram of the Sakaguchi-Kuramoto dynamics.

\subsection{The projected dynamics onto the nodes}
\label{sec:projected}
In order to study Dirac synchronization on any arbitrary network, we proceed using similar steps  {to those} adopted in Ref. \cite{calmon2021topological} for the investigation of Dirac synchronization on a fully connected network. In particular, we use Eq. (\ref{Dirac_synchr}) to determine the dynamics of the phases $\bm\theta$ associated with the nodes of the network and the dynamics of the projection of the link phases onto the nodes of the network, given by $\bm\psi = {\bf K}_0^{-1}{\bf{B}}\bm{\phi}$. In this way we obtain,
\begin{eqnarray}
\dot {\bm {\theta}}&=&\bm\omega -\sigma {\bf K}_0^{-1}{\bf{B}}\sin \left({\bf B}^{\top}(\bm {\theta}+z\bm{\psi})/2)\right),\nonumber \\
\dot {\bm{\psi}}&=&\hat{\bm{\omega}}-\sigma\frac{1}{2} {\bf \hat{L}}_{0}\sin \left(\bm{\psi}-z{\bf \hat{L}}_{0}\bm{\theta}/2
\right),
\label{tpsi}
\end{eqnarray}
where $
\hat{\bm \omega}=\hat{\bf{B}}\tilde{\bm{\omega}}$. Let us  define the $N$-dimensional vectors $\bm\alpha$ and $\bm\beta$ with elements given by 
\bea
\alpha_i&=&(\theta_i+z\psi_i)/2,\nonumber \\
\beta_i&=&z(\theta_i-\Theta_i)/2-\psi_i
\eea
where $\Theta_i$ is given by 
\bea
\Theta_i=\sum_{r=1}^N \frac{a_{ir}}{k_i}\theta_{r}.
\eea
The dynamical Eqs. (\ref{tpsi}) can be expressed elementwise as
\bea
\dot {\theta}_i&=&\omega_i +\sigma \frac{1}{k_i}\sum_{j=1}^N a_{ij}\sin \left(\alpha_j-\alpha_i\right),\nonumber \\
\dot {\psi}_{i}&=&\hat{\omega}_{i}-\sigma \frac{1}{2k_i}\sum_{j=1}^N a_{ij}\left[\sin \left(\beta_j\right)-\sin \left(\beta_i\right)\right].
\label{eq1.}
\eea
Furthermore LDS can be expressed in terms of closed dynamical equations for the of the phases $\bm \alpha$ and $\bm\beta$ (see Appendix $\ref{AppendixB}$).
It follows that the canonical variables of LDS are $\bm\alpha$ and $\bm\beta$ which are linear combinations of nodes and link signals and that the two complex order parameters of the dynamics are given by
\bea
X_{\alpha}=R_{\alpha}e^{\textrm{i}\eta_{\alpha}}=\frac{1}{N}\sum_{j=1}^Ne^{\textrm{i}\alpha_j},\nonumber \\
X_{\beta}=R_{\beta}e^{\textrm{i}\eta_{\beta}}=\frac{1}{N}\sum_{j=1}^Ne^{\textrm{i}\beta_j},
\label{Xab}
\eea
where $R_{\alpha}=|X_{\alpha}|$ and $R_{\beta}=|X_{\beta}|$ indicate the corresponding two real order parameters.


Due to the cross-dimensional coupling introduced by the normalized Dirac operator, a key feature of LDS is that the non-harmonic components of the original signals $\bm \theta$ and $\bm \phi$ are strongly intertwined and the canonical order parameters depend on $\bm \alpha$ and $\bm \beta$, rather than on the  nodes or link signals taken in isolation. 

While this model can be implemented for any distribution of natural frequencies of oscillators $\bf{\omega}$ and $\Tilde{\omega}$ on nodes and links, we draw these frequencies from normal distributions. The frequencies $\omega_i$ have average $\Omega_0$ and precision $\tau_0$, i.e. they are drawn from $\mathcal{N}(\Omega_0,1/\tau_0)$. The choice of the frequencies $\Tilde{\omega}_\ell$ however requires some care when the network is not sparse. Indeed, it is our wish to obtain a set of  projected frequencies $\hat{\bf{\omega}}$ that are themselves normally distributed, with finite variance independent on the network size $N$. This condition is automatically satisfied on sparse networks if the distribution of intrinsic frequencies have a finite variance. However on fully connected networks, as already observed in Ref.~\cite{calmon2021topological,ghorbanchian2020higher}, in order to guarantee that $\hat{\omega}_i$ is normally distributed with mean $0$ and precision $\tau_1$, we need to  require that $\tilde{\omega}_\ell$ is drawn from $\mathcal{N}(\Omega_1,\sqrt{N-1}/\tau_1)$. In the simulation results, unless specified otherwise, the parameters are always taken to be $\Omega_0=\Omega_1=0$, $\tau_0=1$ and $\tau_1 =1$.
\begin{figure}[tbh]
\centering
\includegraphics[width=0.95\columnwidth]{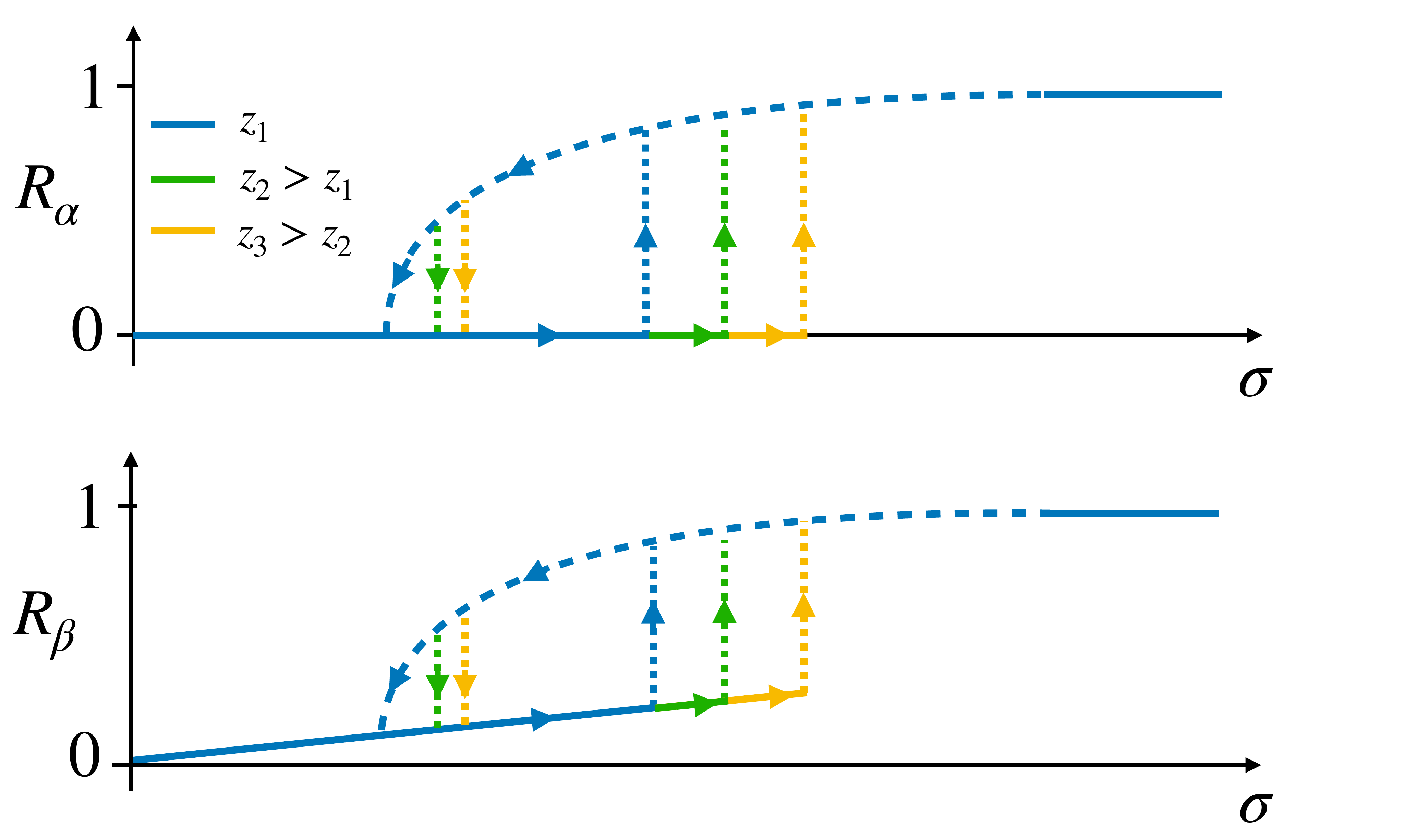}
\caption{The phase diagram of Local Dirac Synchronization (LDS) on sparse networks presents a thermodynamically stable hysteresis loop with a forward discontinuous transition. The san{characteristic} of the backward transition is determined by $z$. As $z$ increases, both transition points occur at larger $\sigma$. Most importantly, the synchronized phase is found to be non-stationary in the complex plane. This rhythmic phase is a key feature of LDS and is a direct consequence of the cross-coupling introduced. Here we represent with a dashed line the region over which this phase can be observed in finite networks. The steady state, shown here in {  solid} line, is found to be a finite size effect. The rhythmic phase is therefore the thermodynamically coherent phase of LDS.
}
\label{fig:transition_schem}      
\end{figure}

\begin{figure}[tbh]
\centering
\includegraphics[width=1\columnwidth]{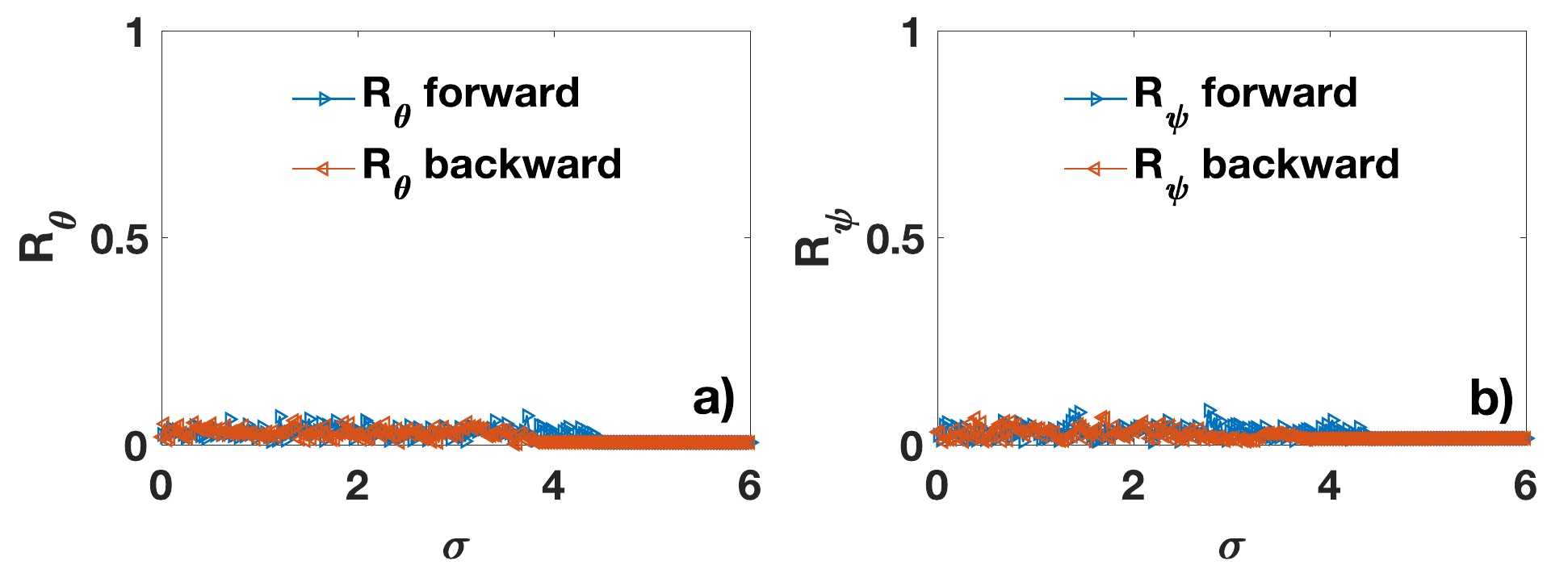}
\caption{{ 
The LDS na\"ive order parameters $R_\theta$ (panel a) and $R_\psi$ (panel b) are shown as a function of the coupling strength   for a Poisson network with $z=10$. It is evident that these observables depending on node and links signals taken in isolation, are not able to capture LDS which strongly couples the dynamics of node and link signals. The Poisson network has $N=800$ nodes and average degree $c=12$. The dynamics is integrated using a $4^\text{th}$ order Runge-Kutta method with integration time $T_{max}=10$, time step $dt = 0.005$, and intervals $\delta\sigma=0.03$. Each order parameter is averaged over the last fifth of the time series at each $\sigma$ step. All results are obtained for $\Omega_0=\Omega_1=0$. } }
\label{fig:naivePoisson}       
\end{figure}

\begin{figure}
\includegraphics[width=1\columnwidth]{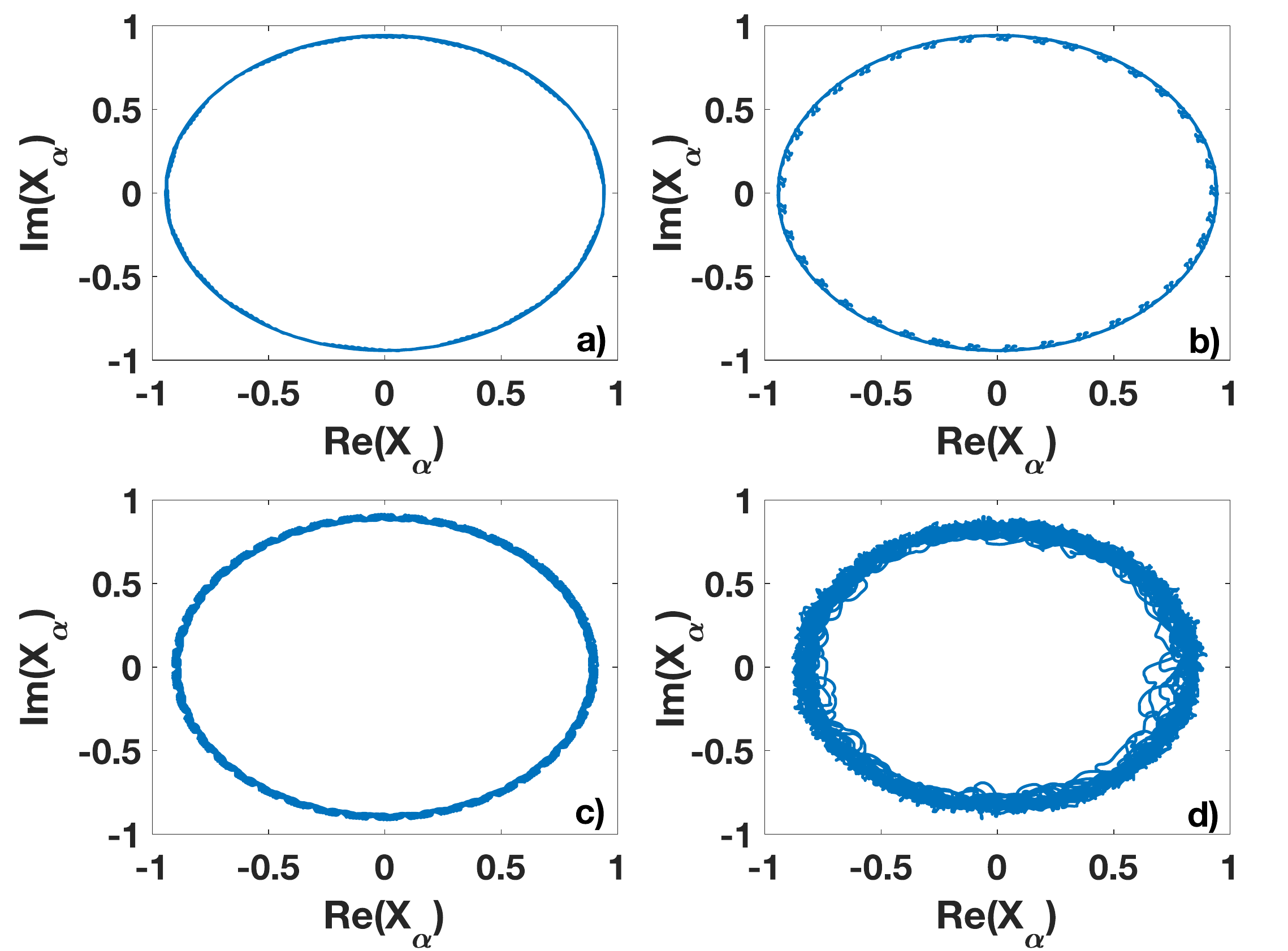}
\caption{The late-time phase portrait of the $X_\alpha$ order parameter is shown over time at $\sigma=3.5$, in the backward transition. The system is left to equilibrate for $T_{max} = 800$ {  and time step $dt = 0.005$} on a Poisson network (panels a and b) and scale-free network (panels c and d) each consisting of $N=100$ nodes. The parameter $z$ was taken to be $4$ in panels a and c and $10$ in panels b and d. All results are obtained for $\Omega_0=\Omega_1=0$.}
\label{fig:portrait_sigma3.5}       
\end{figure}

\begin{figure}
\includegraphics[width=1\columnwidth]{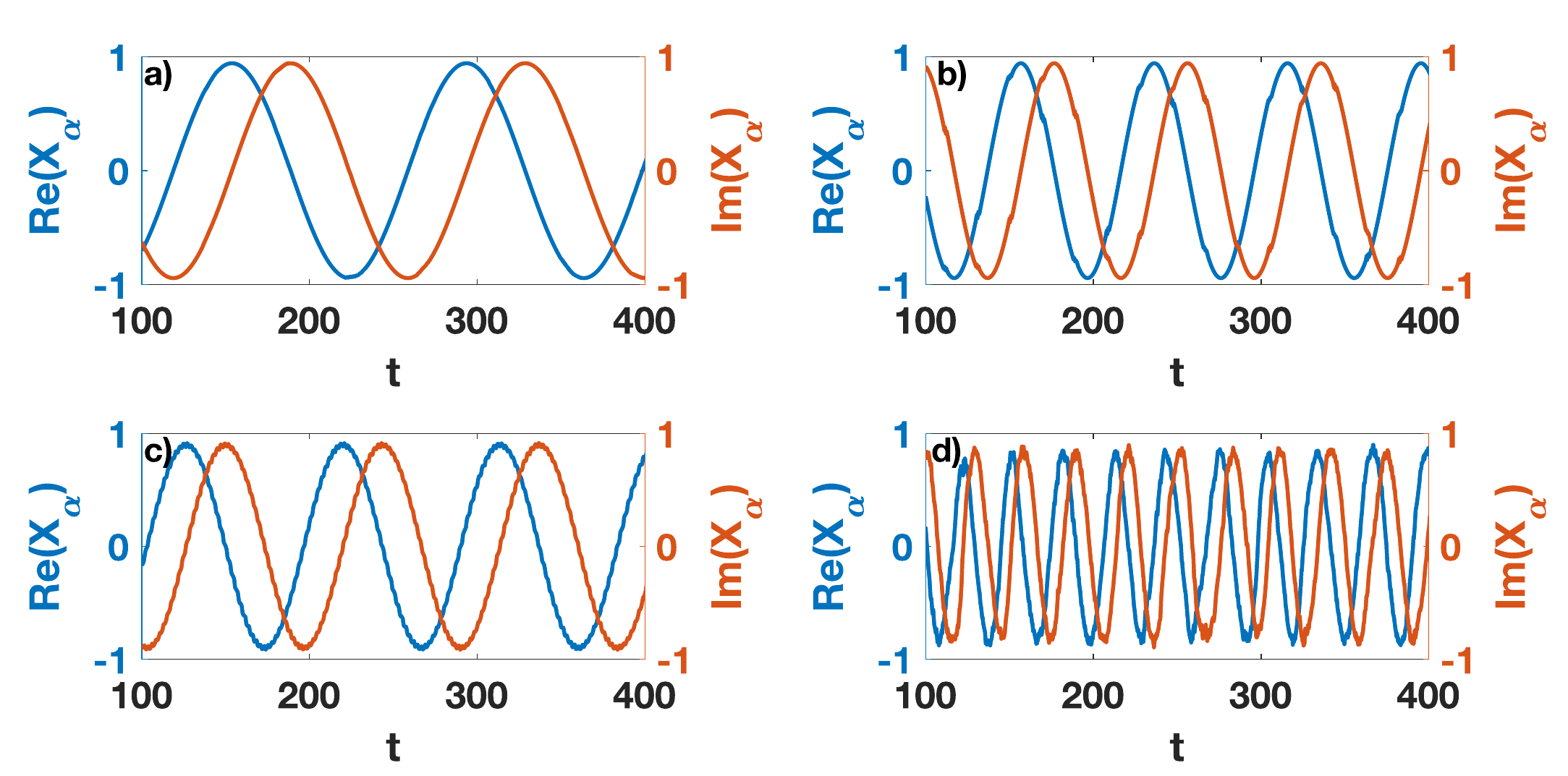}
\caption{{ The time-series of the real and imaginary parts of the  order parameter $X_\alpha$ are shown over time at $\sigma=3.5$ along the backward transition. Panels a and b show results on a Poisson network and panels (c) and (d) show results on a scale-free network. Both networks have $N=100$ nodes. The parameter $z$ has values  $z=4$ in panels a and c and $z=10$ in panels b and d. The time-series are displayed neglecting the transient time. The system is integrated numerically with time step $dt = 0.005$ and $T_{max} = 400$. All results are obtained for $\Omega_0=\Omega_1=0$.}}
\label{fig:time_series_sigma3.5}       
\end{figure}
\section{Explosive synchronization transitions and rhythmic phase} 

Local Dirac Synchronization (LDS) introduces a  non-linear coupling between the phases of the oscillators placed on nodes and links of the network inducing a highly non-trivial feedback mechanism.
In this work, we have investigated  LDS on fully connected and uncorrelated sparse networks such as Poisson and scale-free networks.
From our results we can conclude that the most notable features of LDS are its rich phase diagram and the emergence of a coherent rhythmic phase.
The phase diagram can be captured by the dependence of the real order parameters $R_{\alpha}$ and $R_{\beta}$ on the coupling constant $\sigma$ for any given value of $z$.
{  Here we emphasize that the each of the two  parameters of LDS   ($R_\alpha$ and $R_\beta$)  depends on the dynamical state of both nodes and links, revealing the intertwined dynamics of the two types of topological signals. Indeed the na\"ive order parameters $R_{\theta}$ and $R_{\psi}$ which depend only on nodes and links phases are not able to detect Dirac synchronization in fully connected networks or sparse random networks (see Fig. $\ref{fig:naivePoisson}$ for the  illustrative example of a Poisson network).}
The phase diagram of LDS as a function of $z$ is also very rich  (see schematic representation in Fig.  \ref{fig:transition_schem}) and generalizes the previously obtained phase diagram of Dirac synchronization \cite{calmon2021topological}.
Indeed, for any value of $z$, both order parameters undergo a stable hysteresis loop with a discontinuous forward transition and a backward transition that turns from continuous (for smaller values of $z$) to discontinuous (for larger values of $z$). Moreover, we observe that the size of the hysteresis also grows with $z$.
As we will see in the next section, the resulting phase diagram can be investigated theoretically for fully connected networks. On sparse uncorrelated network, we provide theoretical investigations within the annealed approximation framework.

One of the key features of LDS is that the coherently synchronized phase in which both $X_{\alpha}$ and $X_{\beta}$ are non-zero, (i.e. in finite networks they have values above the finite size fluctuations) is a {\em rhythmic phase}. This rhythmic phase is characterized by the emergence of slow frequency oscillations of the complex order parameter $X_\alpha$ which oscillates at the {\em emergent frequency} $\Omega_E$  in the rotating reference frame in which the intrinsic frequencies have zero mean.

This rhythmic phase has been previously observed for Dirac synchronization on a fully connected network \cite{calmon2021topological}. Here we demonstrate that this phase  persists as a characteristic feature of  LDS as well, as evidenced by the behavior of the model defined on both Poisson and scale-free networks independent of the value of the parameter $z\neq 0$. { This behaviour is represented in Fig. $\ref{fig:portrait_sigma3.5}$ which shows the phase portrait of the complex order parameters $X_{\alpha}$ in the frame rotating with the average intrinsic frequencies.} { The circular patterns observed are characteristic of the rhythmic phase. Indeed, in absence of a rhythmic phase, we would expect the phase portraits of $X_{\alpha}$ to be represented by a single point in the complex plane, and the time-series of the real and imaginary parts of $X_\alpha$ to be stationary. Instead, we observe, that these time-series undergo cyclic oscillations for different values of $z$, on both Poisson and scale-free networks as shown in Fig. $\ref{fig:time_series_sigma3.5}$.} These low frequency oscillations are particularly evident far from the backward  synchronization transition. As the backward synchronization transition threshold is approached, high frequency oscillations of the complex order parameter $X_{\alpha}$ also set in. {  Such phenomena have been observed in the study of brain rhythms \cite{breakspear2010generative,buzsaki2006rhythms}. Thus, despite our system being relatively simple, its ability to capture complex dynamical observations may be harnessed to provide insight into the role of topology as an important element in the onset of brain rhythms.}

As for Dirac synchronization\cite{calmon2021topological}, it can be shown for LDS that the rhythmic phase is the only thermodynamically stable coherent phase in which both order parameters $X_{\alpha}$ and $X_{\beta}$ are non zero.

\begin{figure*}[!tbh]
\centering
\includegraphics[width=1\textwidth]{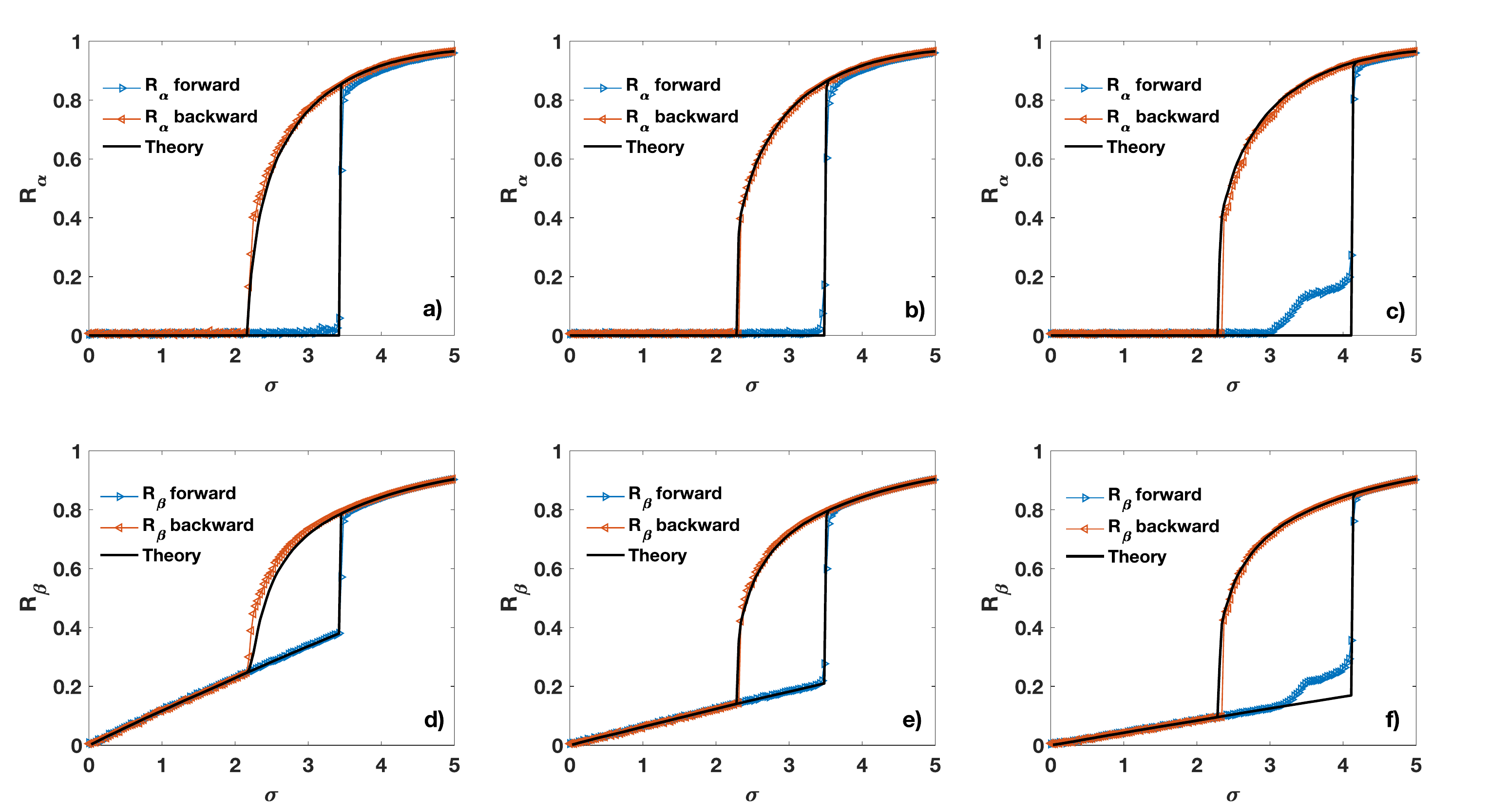}
\caption{The phase diagram of LDS on a fully connected networks is shown by plotting the real order parameters $R_\alpha$ and $R_\beta$  as a function of the coupling constant $\sigma$ for different values of $z$. The values of $z$ are:  $z=5$ (panels a and d), $z=10$ (panels b and e) and $z=15$ (panels c and f). The numerical simulations of the dynamical equations are compared in each case to the analytical solutions. The latter are computed using the numerically observed emergent frequency $\Omega_E$. The theoretical approach fully captures the backward transition, note however that the value of the forward synchronization threshold is numerically obtained. The results are obtained for a network of $N=20000$, integrated using a $4^{\text{th}}$ order Runge-Kutta method with integration time $T_{max}=10$, {  time step $dt = 0.005$} and intervals $\delta\sigma=0.03$. Each order parameter is averaged over the last fifth of the time series at each $\sigma$ step. All results are obtained for $\Omega_0=\Omega_1=0$.
}
\label{fig:transitionFC}       
\end{figure*}
\begin{figure}[tbh]
\centering
\includegraphics[width=1\columnwidth]{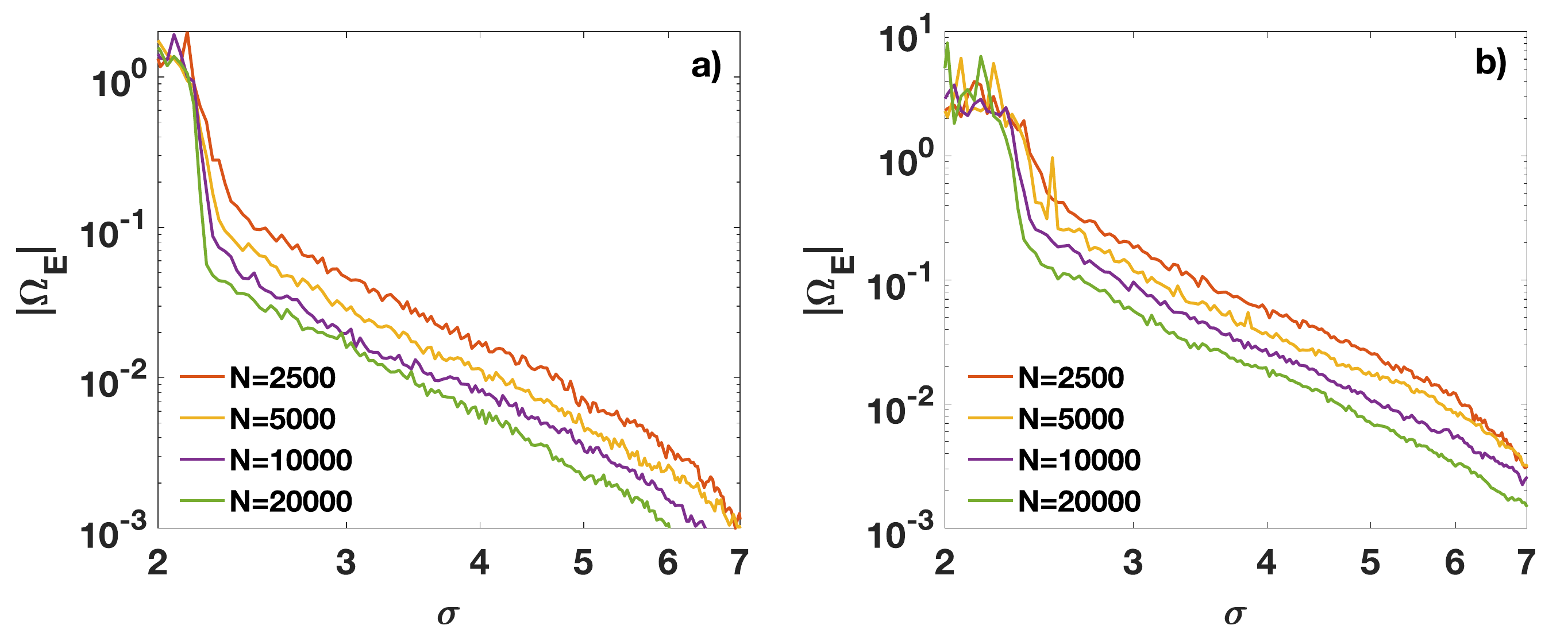}
\caption{The absolute value of the emergent frequency $|\Omega_E|$ of LDS on a fully connected network is plotted versus the coupling constant $\sigma$ for different network sizes. Panels a and b refer to  results obtained for $z=5$  and $z=10$ respectively. Note that the plateau at high values of $\Omega_E$ reached for small values of $\sigma$ corresponds to the incoherent phase.
The dynamical equations on each network are integrated with $T_{max}=10$, $dt = 0.005$ and $\delta\sigma=0.03$, $\Omega_0=\Omega_1=0$.  For each iteration, the emergent frequency measured is averaged over the last fifth of the time series. The absolute value of the emergent frequency $|\Omega_E|$ is averaged over $100$ different iterations of the  backward synchronization transition. All results are obtained for $\Omega_0=\Omega_1=0$.
}
\label{fig:OmegaE_FC}       
\end{figure}
\begin{figure}[!tbh]
\centering
\includegraphics[width=1\columnwidth]{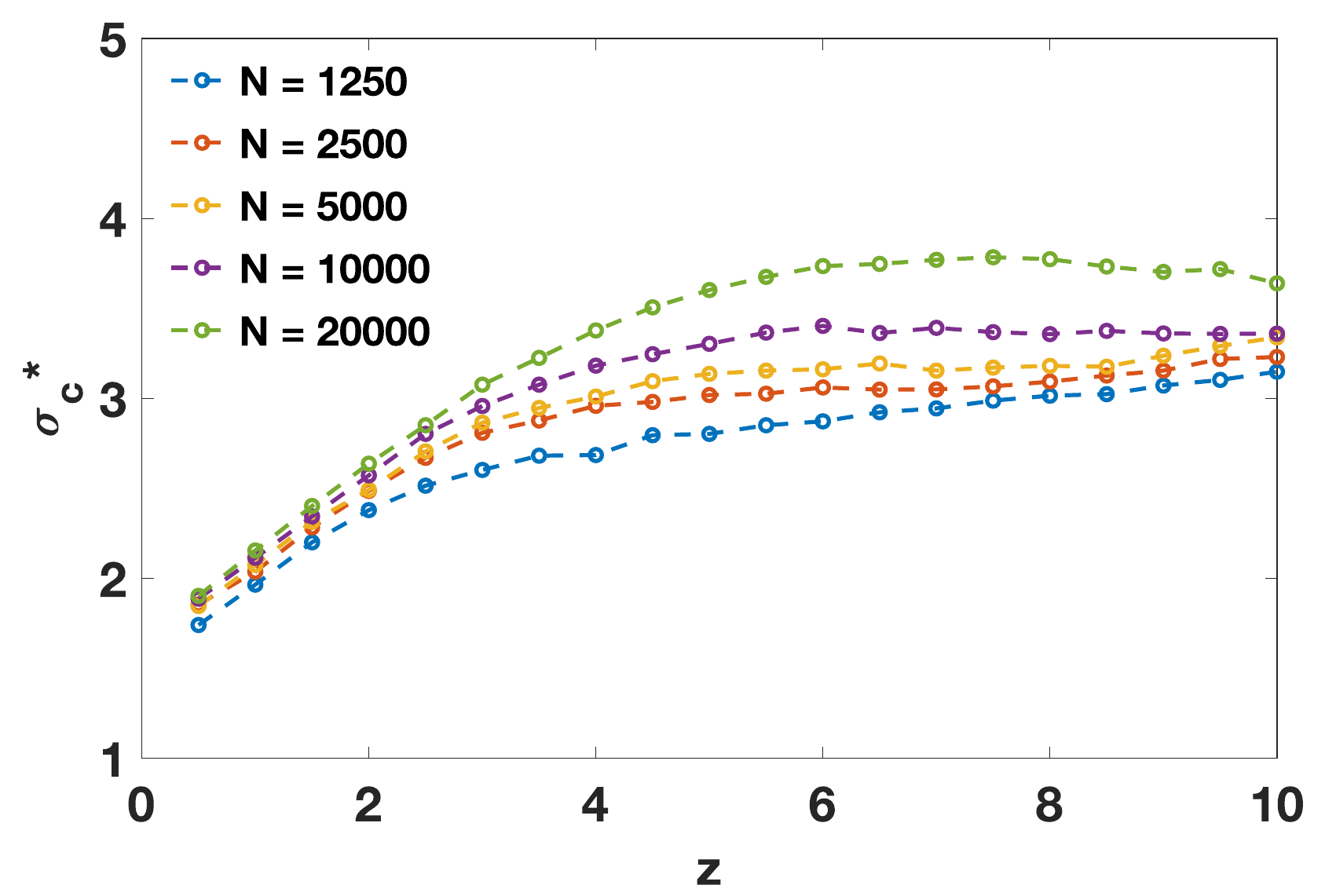}
\caption{The forward synchronization transition $\sigma_c^{\star}$ of the LDS model on a fully connected network is plotted as a function of $z$ for different network sizes $N$. These results have been obtained by numerical integration using  the  integration time $T_{max}=10$, {  time step $dt = 0.005$} and $\delta\sigma=0.03$. Each measurement of $\sigma_c^\star(z)$ is averaged over 25 independent realisations of the phase diagram. All results are obtained for $\Omega_0=\Omega_1=0$.
}
\label{fig:sig_c vs z}       
\end{figure}

Intuitively, we interpret this emergent frequency of LDS as the result of the non-linear dynamics and the non-trivial coupling of chiral modes of the signal, which in the linearized dynamics is responsible for the damped oscillatory behavior of the phases (see Appendix \ref{AppendixA}).
Note that while we can predict the absence of a thermodynamically stable rhythmic phase, the prediction of the value of the  emergent frequency $\Omega_E$ is more challenging and is not an outcome of our analytical treatment.
The value of the emergent frequency $\Omega_E$ can be however measured numerically for networks of different topology and sizes as we discuss in the following.

\section{Phase diagram of Local Dirac Synchronization}
\label{Numerical phase diagram on fully connected and sparse networks}

The phase diagram of Local Dirac Synchronization can be investigated not only numerically but also theoretically. Our theoretical approach  (see Appendix \ref{AppendixC}) provides very good predictions for fully connected networks and for sparse uncorrelated network as well where the latter are obtained  using  the annealed approximation framework.
Interestingly, however  to get accurate predictions both approaches  require as input the knowledge of the (numerically observed) value of the emergent frequency $\Omega_E$.
The numerically obtained phase diagram for the fully connected network  is compared to the theoretical predictions  in Fig. \ref{fig:transitionFC} showing excellent agreement.
The theoretical predictions are obtained using the theoretical approach detailed in the Appendices $\ref{AppendixB}$ and $\ref{AppendixC}$. This approach is similar to the one proposed in Ref. \cite{calmon2021topological} and  uses the numerically observed values of the emergent frequency $\Omega_E$ shown in Fig. \ref{fig:OmegaE_FC}.
Note that the theoretical prediction fully accounts for the backward transition and reveals that this transition, which is continuous for small values of $z$, becomes discontinuous for larger values of $z$, i.e. $z\sim 5$. However, since these predictions make use of the numerically  {measured} values of the emergent frequency $\Omega_E$, we do not have access to a theoretical prediction of the critical value of $z$ at which the backward transition becomes discontinuous.
The forward transition is always discontinuous in LDS and may be predicted using similar methods developed in in Ref. \cite{calmon2021topological}. Note however that the forward synchronization threshold $\sigma_c^{\star}$ of LDS is strongly affected by finite size effects. These finite size effects  are responsible for the onset of the instability of the incoherent phase for values of the coupling constant that are lower than the forward synchronization threshold in infinite networks. In order to study numerically the dependence of these finite size effects on the value of $z$ we plot  {in Fig. \ref{fig:sig_c vs z}}
$\sigma_c^{\star}$ as a function of $z$ for different network sizes  {and show} that $\sigma_c^{\star}$ increases with $z$ in the thermodynamic limit. { However} for finite networks we observe a plateau of $\sigma_c^{\star}$ as a function of $z$.

\begin{figure}[tbh]
\centering
\includegraphics[width=1\columnwidth]{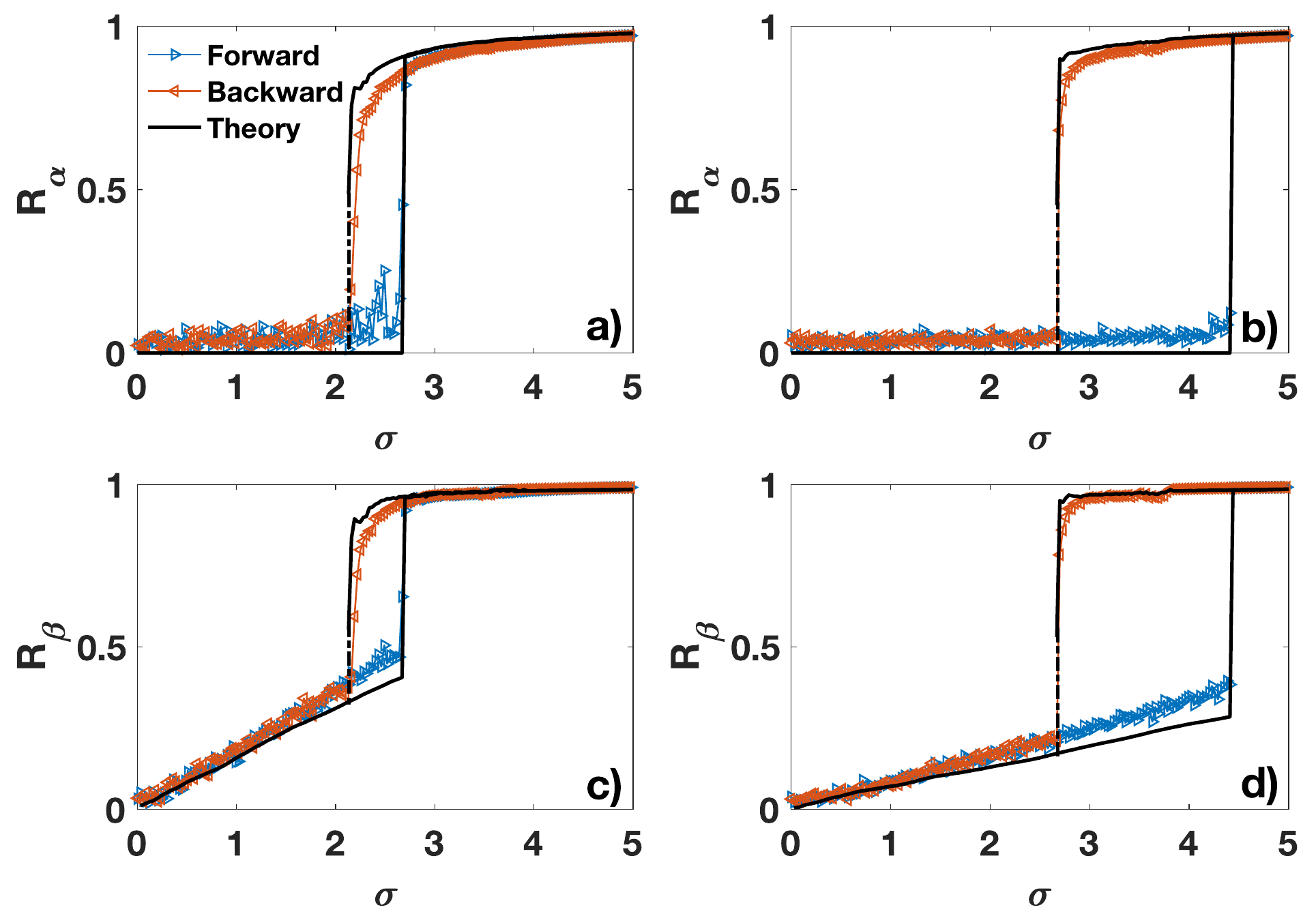}
\caption{
The phase diagram of LDS on a Poisson network is shown for different values of $z$ and compared to the theoretical predictions obtained in the annealed approximation. Panels a and c display the  parameters  $R_\alpha$ and $R_\beta$ as a function of the coupling constant $\sigma$ for $z=4$, panels b and d correspond instead to $z=10$.  The theoretical predictions are computed using the numerically observed emergent frequency $\Omega_E$ and degree distribution of the network. The position of the forward transition is affected by strong finite size effects and is not predicted by the theory. The numerical results are obtained for a Poisson network of $N=800$, with average degree $c=12$, integrated using a $4^\text{th}$ order Runge-Kutta method with integration time $T_{max}=10$, {  time step $dt = 0.005$}, and intervals $\delta\sigma=0.03$. Each order parameter is averaged over the last fifth of the time series at each $\sigma$ step. All results are obtained for $\Omega_0=\Omega_1=0$. }
\label{fig:transitionPoisson}       
\end{figure}

\begin{figure}[tbh]
\centering
\includegraphics[width=1\columnwidth]{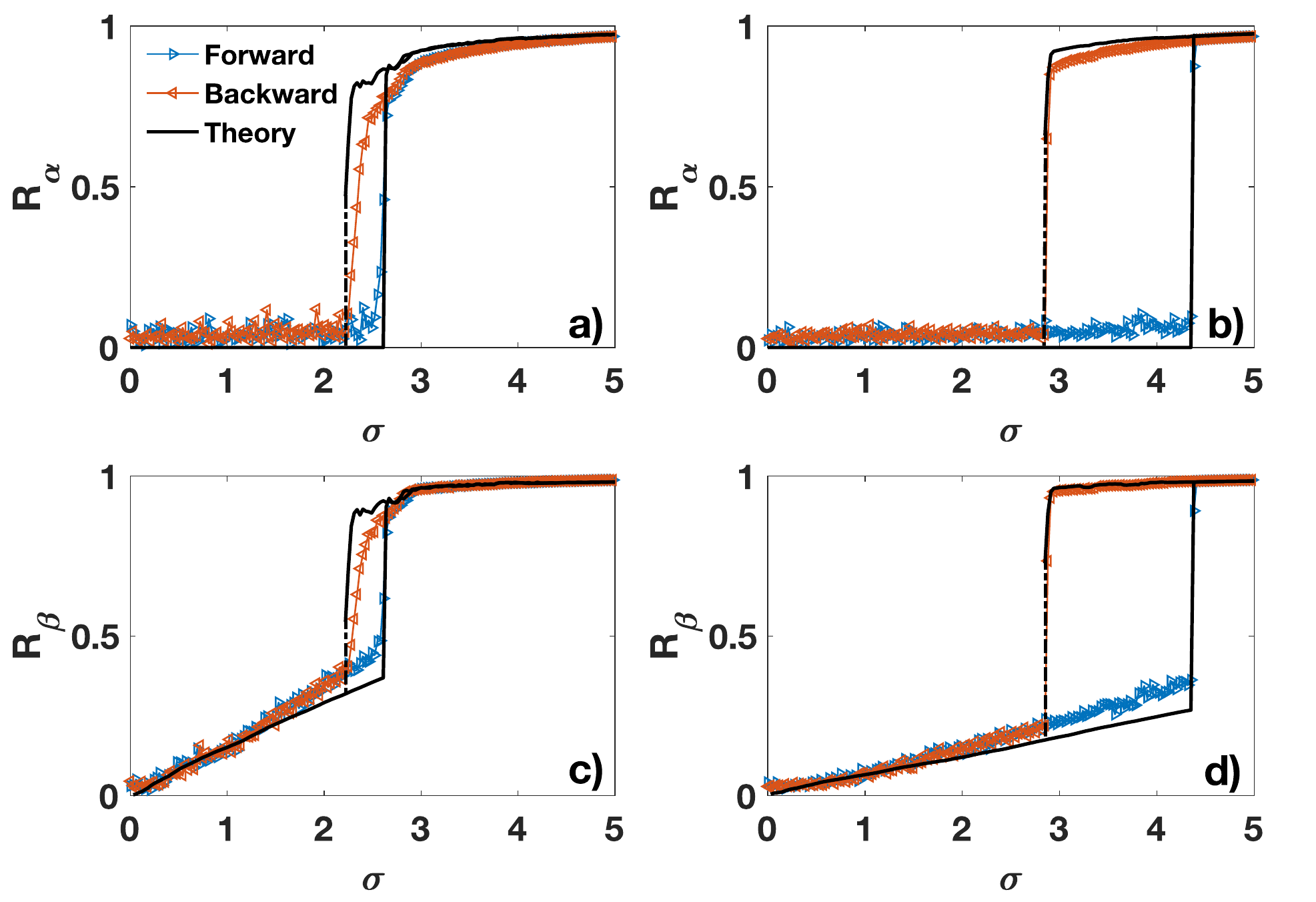}
\caption{The phase diagram of LDS on a scale-free network network is shown for different values of $z$ and compared to the theoretical predictions obtained in the annealed approximation. Panels a and c display the  parameters  $R_\alpha$ and $R_\beta$ as a function of the coupling constant $\sigma$ for $z=4$, panels b and d correspond instead to    $z=10$.  The theoretical predictions are computed using the numerically observed emergent frequency $\Omega_E$ and degree distribution of the network. The position of the forward transition is affected by strong finite size effects and is not predicted by the theory. The numerical results are obtained for a scale-free network of of $N=800$ nodes with minimum degree $6$, structural cutoff and power-law exponent  $\tilde{\gamma}=2.5$. The dynamical equations are  integrated using a $4^\text{th}$ order Runge-Kutta method with $T_{max}=10$, {  time step $dt = 0.005$}, and intervals $\delta\sigma=0.03$. Each order parameter is averaged over the last fifth of the time series at each $\sigma$ step. All results are obtained for $\Omega_0=\Omega_1=0$.}
\label{fig:transitionSF}       
\end{figure}

For sparse uncorrelated networks such as Poisson and scale free networks, theoretical estimations to the phase diagrams can be obtained using the annealed approximation as detailed in Appendices $\ref{AppendixB}$ and $\ref{AppendixC}$. 
The numerical results on both Poisson (Fig. \ref{fig:transitionPoisson}) and scale-free networks (Fig.  \ref{fig:transitionSF}) are in good agreement with the theoretical results despite the fact that the annealed approximation essentially neglects the effect of the locality of the interactions between nodes and links which is an essential element of LDS. 
{  Note that the annealed approximation can be used to study Dirac synchronization on random uncorrelated networks with arbitrary degree distribution but it cannot be used to study the model on an arbitrary network topology. For instance for networks with a finite spectral dimensions, different approaches along the ones adopted in Ref. \cite{millan2019synchronization} should be considered.} 

\begin{figure}[!tbh]
\includegraphics[width=1\columnwidth]{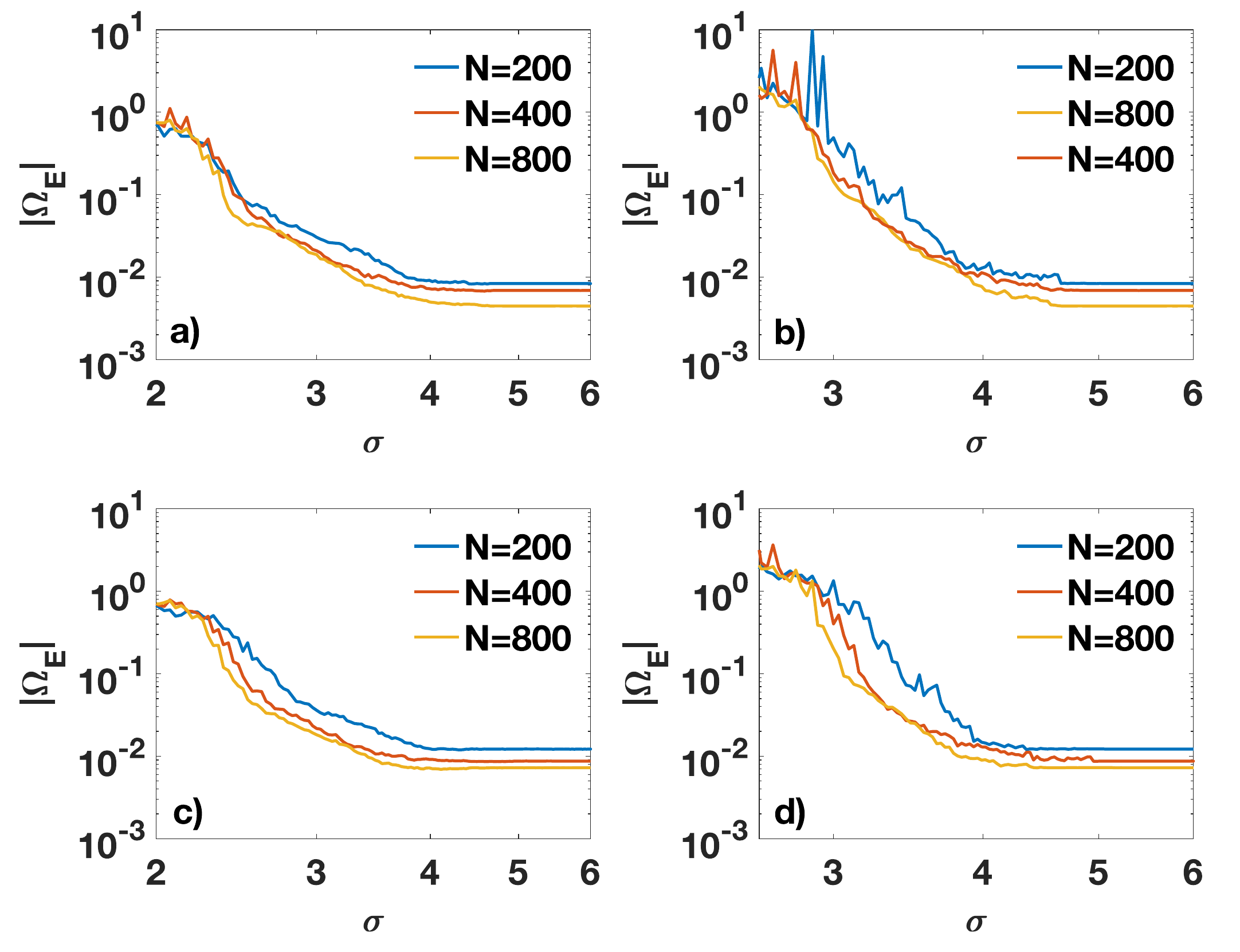}
\caption{The absolute value of the emergent frequency $|\Omega_E|$, monitored across the backward transition,  is shown as a function of the coupling constant $\sigma$ for a Poisson network with average degree $c=12$ (panels a and b) and a scale free network of minimum degree $6$, structural cutoff, and power-law exponenet $\tilde{\gamma}=2.5$ (panels c and d). The networks have  various system sizes $N$.  The value of $z$ are:  $z=4$ (panels a and c) and $z=10$ (panels b and d). For each network size, the absolute value of the emergent frequency is averaged over $100$ independent initial conditions and natural frequencies. All results are obtained for $\Omega_0=\Omega_1=0$.}
\label{fig:emergentSFRG}       
\end{figure}
\begin{figure}
\includegraphics[width=1\columnwidth]{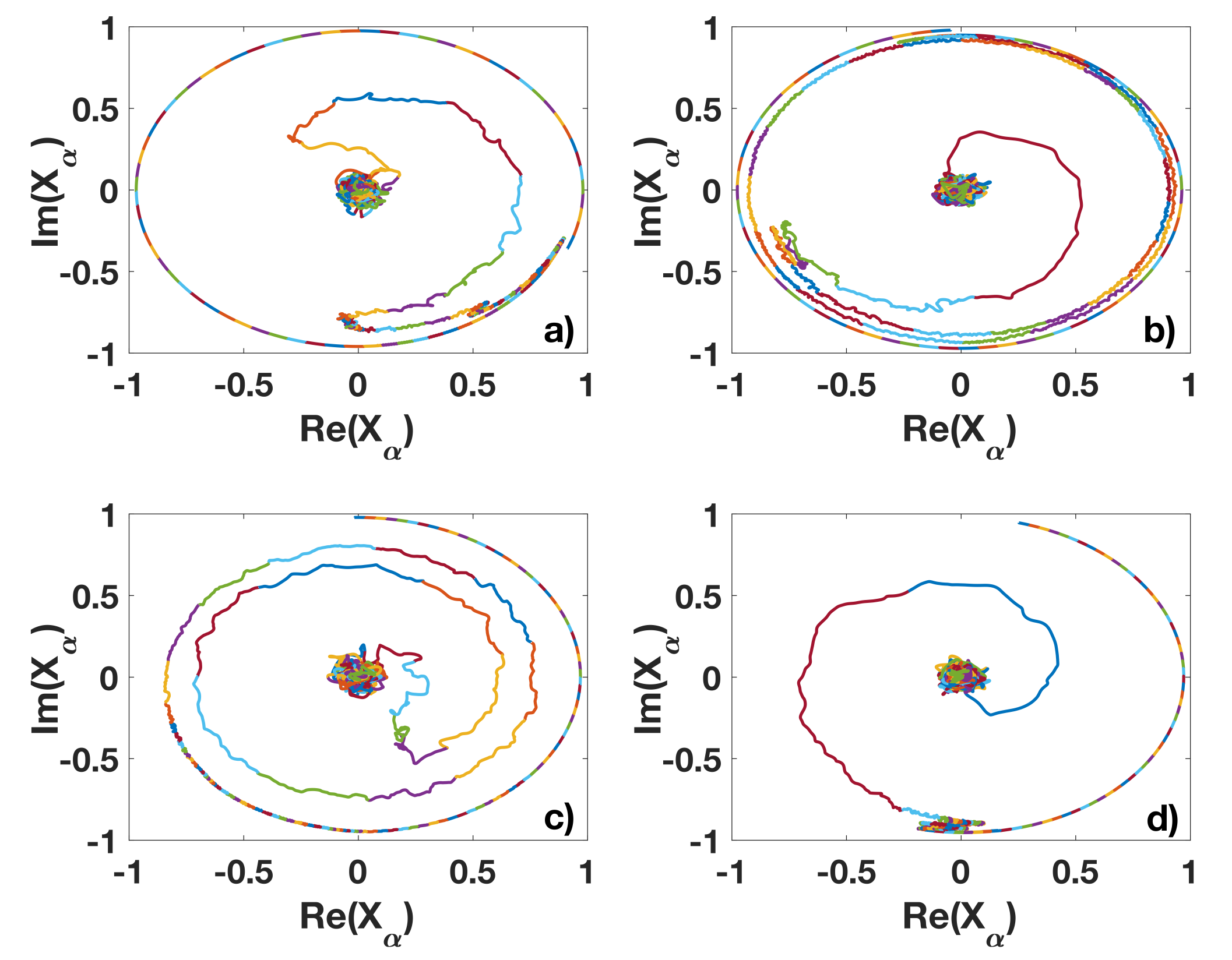}
\caption{The phase portrait of the complex order parameter  $X_\alpha$ for a Poisson (panels a and b) and scale-free network (panels c and d) is shown over the backward transition. Each line of different color corresponds to distinct $\sigma$ steps, during which the system equilibrates over the integration time $T_{max} = 10$ {  and time step $dt = 0.005$}. Simulations are obtained on a Poisson network with average degree $c=12$ (panels a and b) and on a scale-free  network (panels c and d) with minimum degree $6$, structural cutoff and power-law exponent $\tilde{\gamma}=2.5$. Both networks have $N=800$ nodes. The value of the  parameter $z$ is  $z=4$ in panels a and c and $z=10$ in panels b and d. All results are obtained for $\Omega_0=\Omega_1=0$.}
\label{fig:portrait}       
\end{figure}

\begin{figure}
\includegraphics[width=1\columnwidth]{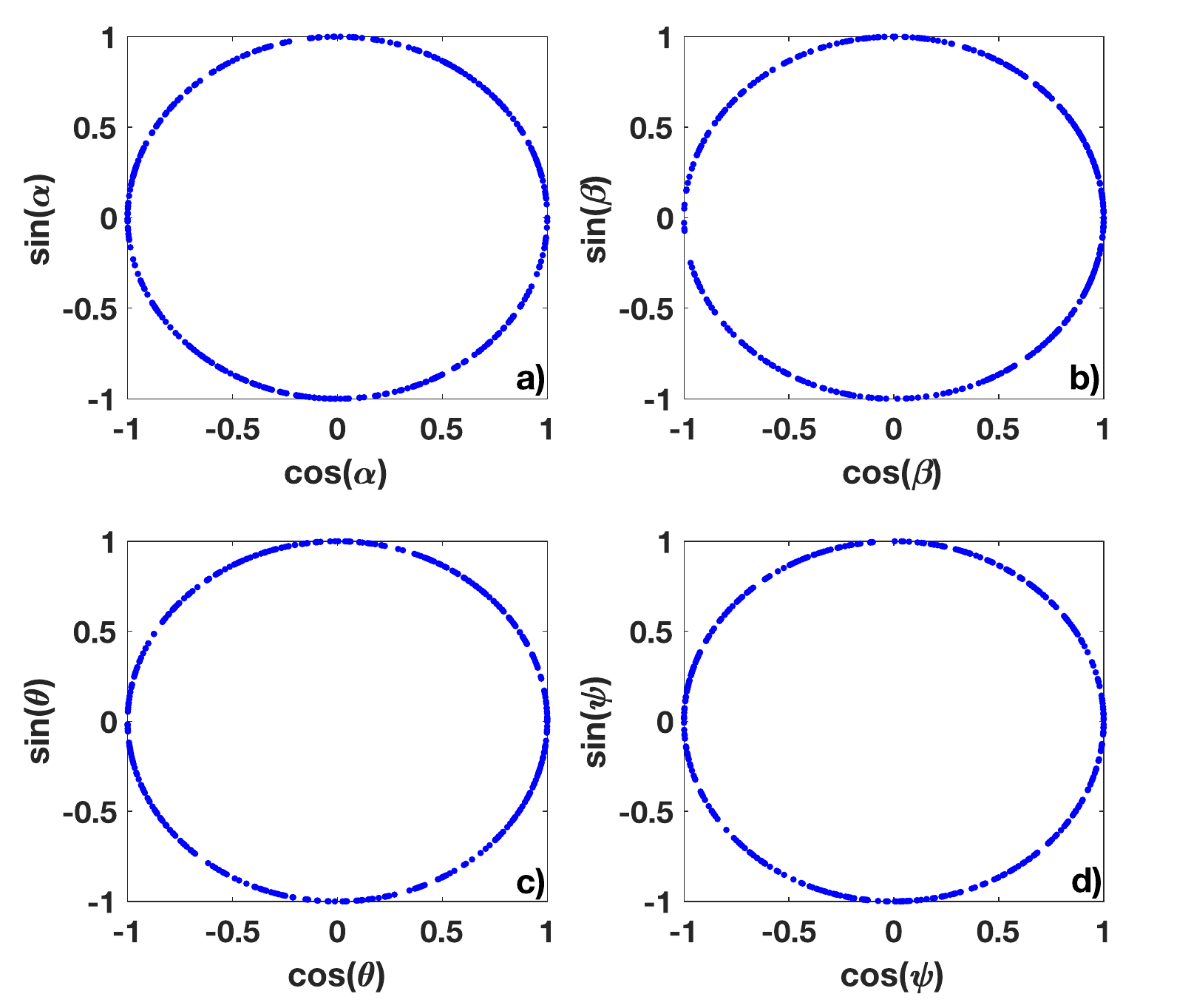}
\caption{{ A snapshot of the phases $\alpha$, $\beta$, $\theta$ and $\psi$ is shown for a Poisson network with average degree $c=12$ in the incoherent phase. These results were obtained by numerical integration using a $4^\text{th}$ order Runge-Kutta method with integration time $T_{max}=10$ and time step $dt = 0.005$, at fixed coupling strength $\sigma = 1$ and $z=4$. All results are obtained for $\Omega_0=\Omega_1=0$.}}
\label{fig:snapshot_sigma1}       
\end{figure}

\begin{figure}
\includegraphics[width=1\columnwidth]{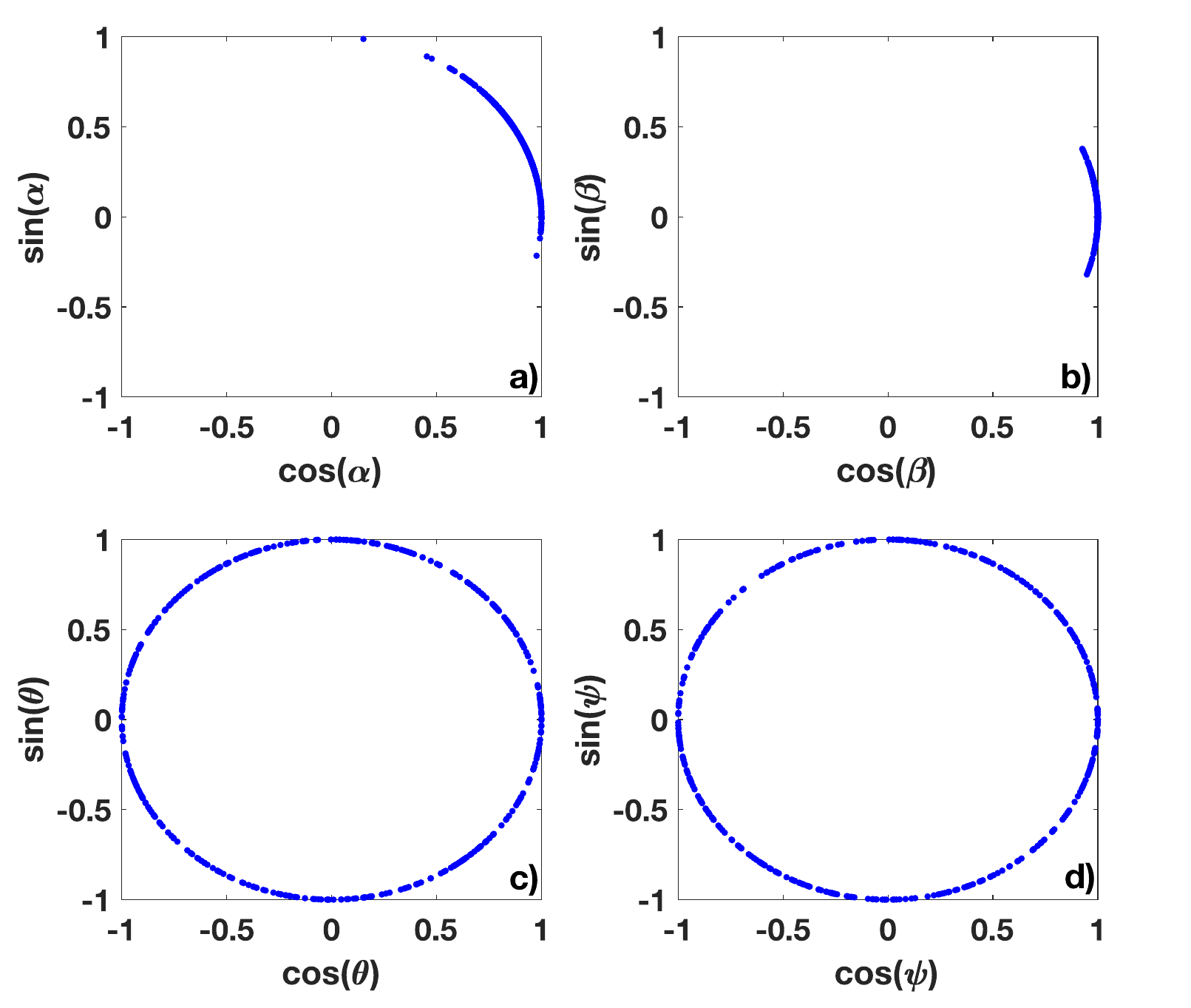}
\caption{{ A snapshot of the phases $\alpha$, $\beta$, $\theta$ and $\psi$ is shown for a Poisson network with average degree $c=12$ in the synchronized phase. These results were obtained by numerical integration using a $4^\text{th}$ order Runge-Kutta method with integration time $T_{max}=10$ and time step $dt = 0.005$, at fixed coupling strength $\sigma = 5$ and $z=4$. All results are obtained for $\Omega_0=\Omega_1=0$.}}
\label{fig:snapshot_sigma5}       
\end{figure}
For sparse Poisson and scale-free networks,  similar to what occurs for fully connected networks, the backward transition is  influenced by the value of $z$. For smaller value of $z$  (such as $z=4$ in Figs. \ref{fig:transitionPoisson} and $\ref{fig:transitionSF}$), the backward transition is continuous on both scale free and Poisson networks. However for larger values of $z$ (such as $z=10$ as shown in \ref{fig:transitionPoisson} and \ref{fig:transitionSF}) this transition becomes discontinuous. As revealed previously, the coherent phase is  a rhythmic phase with emergent frequency $\Omega_E$ (see Fig. $\ref{fig:emergentSFRG}$). We note that this emergent frequency  decreases with larger network sizes {  and eventually plateaus} for sparse Poisson and scale-free networks (see Fig. \ref{fig:emergentSFRG}) as {  while for  fully connected networks, it monotonically decreases }(see Fig. $\ref{fig:OmegaE_FC}$).
Finally the desynchronization transition of the LDS can also be visualized by plotting the phase portrait  of the complex order parameter $X_{\alpha}$ (see Fig. \ref{fig:portrait}) in the rotating frame of the intrinsic frequencies.  
{ In order to provide a microscopic description of the transition in Figs. \ref{fig:snapshot_sigma1} and $\ref{fig:snapshot_sigma5}$ we display the typical snapshots of the phases of the network mapped on a unit circle  for dynamical configurations below and above the synchronization transition. We observe that the transition is captured by the distribution of phases $\bm \alpha$ and $\bm\beta$ which clearly clusters above the transition. In contrast, the distribution of the phases $\bm\theta$ and $\bm\psi$ remains fairly homogeneous on the unit circle below as well as above the transition, indicating that LDS is only revealed by the phases $\bm\alpha$ and $\bm \beta$ that are linear combination of node and link signals. This is a further confirmation that the phases $\theta$ and $\psi$ are strongly correlated, and when taken in isolation, are unable to describe LDS. By looking at the dynamics of the phase snapshots for non-zero $\Omega_0$ and $\Omega_1$ (not shown here) we also observe that, as expected, the phases $\bm \theta$ and $\bm \alpha$ drift due a combined effect of non-zero average intrinsic frequency and the emergence of the rhythmic phase, while the phases $\bm\psi$ and $\bm\beta$ do not appear to drift. This phenomenon is related to a fundamental difference between these phases: while $\bm\theta$ and $\bm\alpha$ capture the harmonic modes of the phases associated with the nodes, $\bm \psi$ and the derived phases $\bm\beta$ are constructed from links phases projected onto the nodes by the coboundary operator, therefore the harmonic component which encodes the non trivial synchronized dynamics is filtered out.}

Overall, these results  confirm that LDS can be naturally defined on sparse 
networks using the normalized Dirac operator, and reveal that LDS on sparse networks displays a phenomenology which is in line with the one of Dirac synchronization defined on fully connected networks in Ref. \cite{calmon2021topological}.

\begin{figure}
\includegraphics[width=1\columnwidth]{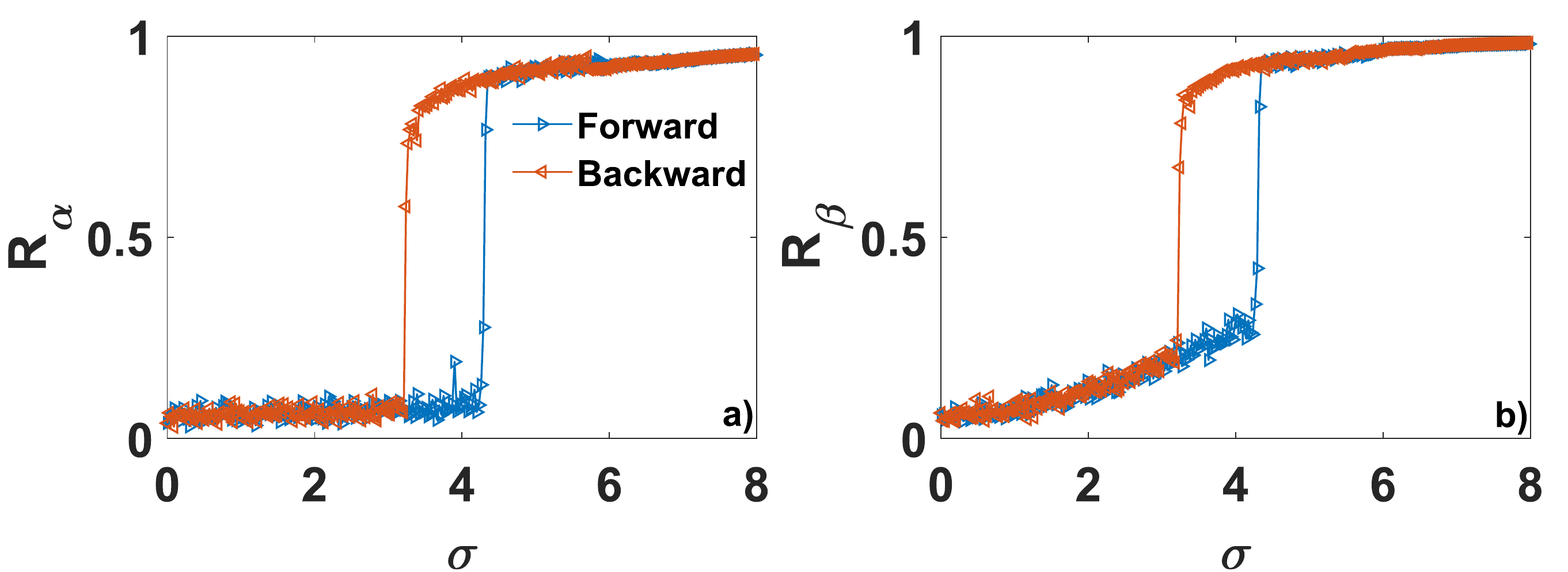}
\caption{{ The phase diagram of LDS defined on the synaptic monadic layer ("MonoSyn"/layer 2) of the C-elegans multiplex connectome \cite{uno,due,Manlio_repository}  is shown for $z=12$. Panels a and b display the  parameters  $R_\alpha$ and $R_\beta$ as a function of the coupling constant $\sigma$ for $z=12$. The results are obtained by numerical integration using a $4^\text{th}$ order Runge-Kutta method with integration time $T_{max}=10$, time step $dt = 0.0005$ and intervals $\delta\sigma=0.03$. Each order parameter is averaged over the last fifth of the time series at each $\sigma$ step. All results are obtained for $\Omega_0=\Omega_1=0$.}}
\label{fig:c-elegans-phase}       
\end{figure}

\begin{figure}
\includegraphics[width=1\columnwidth]{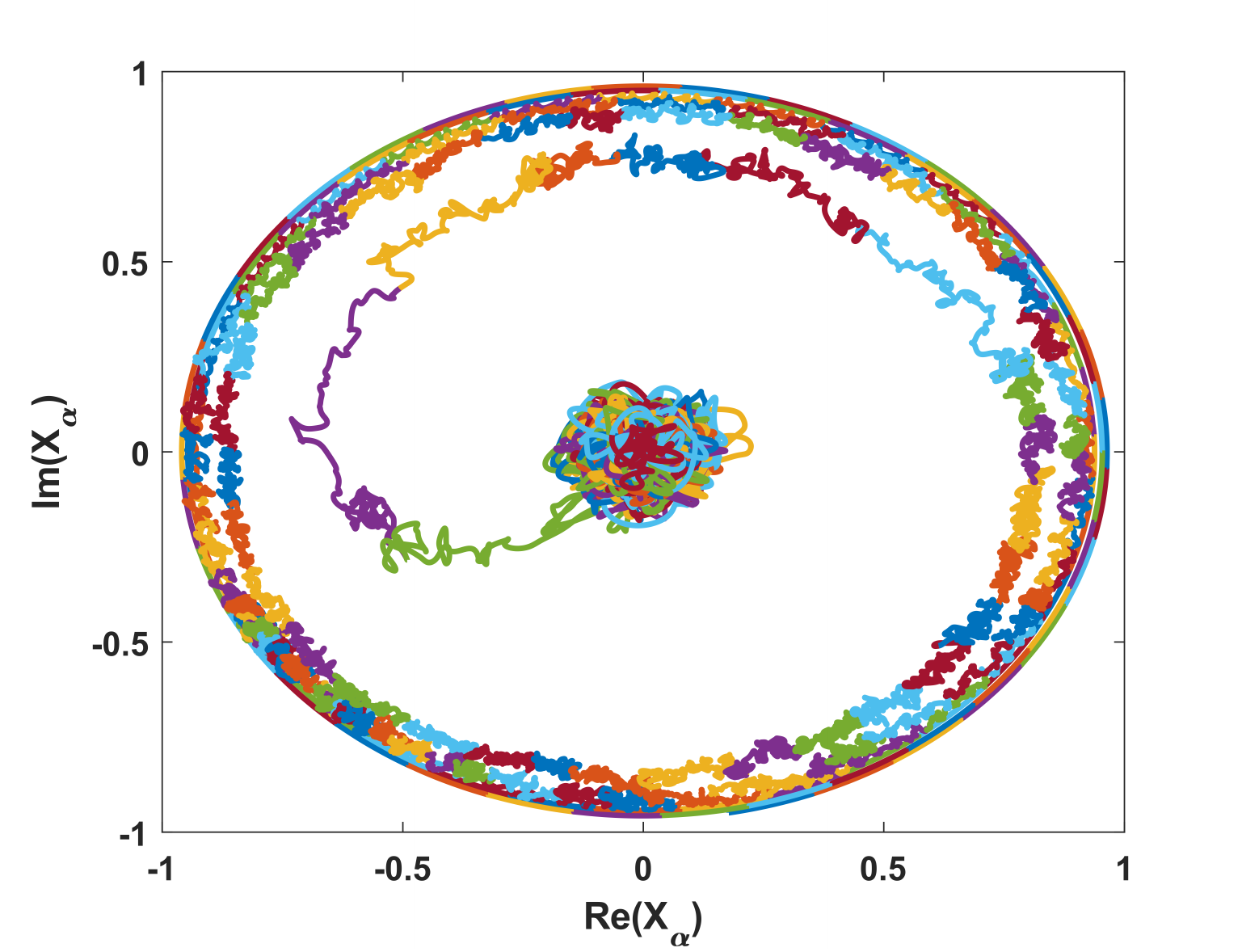}
\caption{{ The phase portrait of the complex order parameter $X_\alpha$ over the backward transition for LDS defined on the synaptic monadic layer ("MonoSyn"/layer 2) of the C-elegans multiplex connectome \cite{uno,due,Manlio_repository} reveals the rhythmic phase. Each line of different color corresponds to distinct $\sigma$ steps, during which the system equilibrates over the integration time $T_{max} = 10$ and time step $dt = 0.0005$. All results are obtained for $\Omega_0=\Omega_1=0$.}}
\label{fig:c-elegans-portrait}       
\end{figure}

{  So far we have discussed LDS on fully connected and on sparse random graphs. This allows us to compare the numerical results with theoretical predictions derived in the annealed approximation. An open question that emerges is whether these results apply to more general topologies. In particular, it remains to investigate whether the rhythmic modes that are observed in brain systems can be captured by our framework considered on real brain networks. In order to address this question we have numerically investigated LDS on the synaptic layer of the neuronal network of Caenorhabditis elegans (C.elegans) \cite{uno,due}.
We observe results that are qualitatively in line with the results obtained on uncorrelated sparse networks. In particular, the phase diagram of LDS displays an hysteresis loop with a sharp forward transition and a backward transition that becomes abrupt for larger value of the parameter $z$ (see Fig. $\ref{fig:c-elegans-phase}$). Additionally, the phase portrait of the order parameter $X_{\alpha}$ across the desynchronization transition reveals that the rhythmic phase emerges for LDS defined on the connectome of C.elegans (see Fig. \ref{fig:c-elegans-portrait}).}

\section{Conclusions}
In this paper, we formulate Local Dirac Synchronization (LDS), a dynamical model that uses the normalized Dirac operator to describe the coupled dynamics of oscillators placed on both nodes and links of an arbitrary connected network.
This model couples node and link signals by introducing, in the dynamics of the nodes oscillators, a phase lag depending of the phases of nearby links, and vice versa, introducing in the dynamics of link oscillators a phase lag depending of the phases of nearby nodes. In both cases the magnitude of the phase-lag is parametrized by the parameter $z$. Hence the model is inspired by the Sagakuchi-Kuramoto model but differs from it as the phase-lag is not constant but rather adaptive. 
Local Dirac synchronization reveals several phenomena that emerge from the rich interplay between network topology and dynamics.
First, the collective motion includes the free oscillations of the harmonic mode(s) of the dynamics.  This phenomenon reveals the important dynamical effects caused by the presence of cavities and non-zero Betti numbers in the network structure. Second, the dynamics of the nodes and links oscillators are intertwined as  the order parameters  are given by linear combinations of the nodes and link signals. Third, LDS displays the emergence of a collective rhythmic phase in which one of the complex order parameters oscillates  in the reference frame  that rotates together with the intrinsic frequencies. This phenomenon can reveal how topology could be responsible for the emergence of brain rhythms. 
Fourth, the phase diagram of the model is informative and displays a discontinuous forward transition and a backward transition which is either continuous or discontinuous depending on choice of the parameter $z$.

Thus, in this paper we have formulated Local Dirac Synchronization, motivating the model choice and comparing its phenomenology with the recently introduced Topological Kuramoto model. The resulting phenomenology is studied numerically and analytically on both Poisson and scale-free networks. The numerical results on fully connected networks are in excellent agreement with the theory, the results on sparse networks are well approximated by the theoretical predictions obtained within the annealed approximation.
{ Finally we provide numerical evidence that our results derived for sparse uncorrelated networks are in strong qualitative agreement with the phenomenology of LDS defined on the real topology of the synaptic layer of C.elegans.}

These results open new perspectives in the  dynamics of coupled topological signals on networks and simplicial complexes which are promising directions to investigate the interplay between topology and dynamics with possible applications to brain research amongst other applications.

\begin{acknowledgments}
This research utilised Queen Mary's Apocrita HPC facility, supported by QMUL Research-IT. http://doi.org/10.5281/zenodo.438045
G.B. acknowledges support from the Royal Society (IEC\textbackslash NSFC\textbackslash191147 and The Alan Turing Institute.
\end{acknowledgments}
\section*{Conflict of Interest Disclosure Statements}
The authors have no conflicts to disclose.

\section*{Data Availability Statement}
The data of the neural network of Caenorhabditis elegans used in this study is formed by the synaptic monadic layer ("MonoSyn/layer 2) of the C.elegans multiplex connectome \cite{uno,due} publicly available at the repository \cite{Manlio_repository}.

\section*{Author contribution}
G.B. designed the research; L.C., S. K. and G.B. conducted the research and wrote the manuscript.

\bibliography{references}

\appendix
\section{Linearized dynamics of the topological Kuramoto model and of the Local Dirac synchronization}
\label{AppendixA}
We consider here the linearized dynamics of the topological Kuramoto model and compare it to the linearized dynamics of Local Dirac Synchronization.
 
The linearized dynamics of the topological Kuramoto model expressed in terms of the normalized Dirac operator is given by  Eq. (\ref{TKlin_unnorm}) that we rewrite here for convenience,
\bea
\dot{\bm \Phi}=\bm\Omega-\sigma {\hat{\mathcal{L}}}{\bm \Phi}.
\eea
Let us  decompose the dynamical vector $\bm\Phi$ and the vector of the intrinsic frequencies $\bm\Omega$ into the eigenvectors $\Psi_{\lambda}$ with eigenvalue $\lambda$ of the normalized Dirac operator ${\bf \hat{D}}$, i.e. 
\bea
\bm\Phi=\sum_{\lambda}c_{\lambda}\bm{\Psi}_{\lambda}.\nonumber \\
\bm \Omega=\sum_{\lambda}\Omega_{\lambda}\bm{\Psi}_{\lambda}.
\label{decomposition}
\eea
Note that here  we consider a  notation in which we do not specify whether the vectors $\bm\Psi_{\lambda}$ are the left or right eigenvectors of the normalized Dirac operator since the following derivation is independent  {of} the choice of basis adopted.
The linearized equation can then  {be} expressed as 
\bea
\dot{c}_{\lambda}=\Omega_{\lambda}-\sigma \lambda^2c_{\lambda},
\eea
 {which admits the solution}
\bea
c_{\lambda}(t)=\frac{\Omega_{\lambda}}{\sigma\lambda^2}\left(1-e^{-\sigma \lambda^2 t}\right)+c_{\lambda}(0)e^{-\sigma \lambda^2 t}.
\eea
as long as $\lambda\neq 0${. Instead, }for the harmonic components ($\lambda=0$) we obtain
\bea
c_{harm}(t)=\Omega_{harm}t+c_{harm}(0).
\eea
The harmonic modes {therefore} oscillate at a frequency dictated by the intrinsic frequencies, while the non-harmonic modes are damped and freeze asymptotically in time.
Note that since for every eigenvalue $\lambda^2\neq 0$ of the super-Laplacian $\mathcal{\hat{L}}$ the normalized Dirac operator has one positive and one negative eigenvalue given by $\pm \lambda$, and since the corresponding  eigenvectors are related by chirality, the linearized dynamics can be decomposed into two independent dynamics for the nodes and for the link signals.
The generalization of this derivation necessary to treat the linearized  dynamics dictated by the un-normalized Dirac operator given by Eq. (\ref{TKlin_norm}) is straightforward.

Let us now compare the linearized dynamics of the Topological Dirac Kuramoto model with the linearized dynamics of the Local Dirac Synchronization which is given by Eq.~(\ref{Dirac_linearised}) that we rewrite here for convenience, 
\bea
\dot{\bm\Phi}=\bm\Omega-\sigma ({\bf \hat{D}^2}+z\bm\gamma{\bf \hat{D}}^3)\bm \Phi,
\eea
where $\bm\gamma$ is given by Eq.~(\ref{gamma}).
By decomposing $\bm\Phi$ and $\bm\Omega$ into the left eigenvectors $\bm\Psi_{\lambda}$
of the normalized Dirac operator as in Eq.~(\ref{decomposition}) using the chirality of the eigenvectors of the normalized Dirac operator, we can express the linearized dynamics as 
\bea
\begin{array}{lll}
&\dot{c}_{\lambda}=\Omega_{\lambda}-\sigma(\lambda^2c_{\lambda}-z\lambda^3c_{-\lambda}) &\mbox{for}\ \lambda\neq 0,\\
&\dot{c}_{harm}=\Omega_{harm} & \mbox{for}\ \lambda=0.
\end{array}
\eea
Which for $\lambda>0$ has solution 
\bea
\begin{pmatrix}c_{\lambda}(t)\\c_{-\lambda}(t)\end{pmatrix}=A(t)\begin{pmatrix}1\\-\textrm{i}\end{pmatrix}+B(t)\begin{pmatrix}1\\\textrm{i}\end{pmatrix},
\label{uno}
\eea
where 
\bea
A(t)=\frac{\Omega_{\lambda}+\textrm{i}\Omega_{-\lambda}}{2\sigma(\lambda^2+\textrm{i}z\lambda^3)}\left(1-e^{-\sigma(\lambda^2+\textrm{i}z\lambda^3)t}\right)+A(0)e^{-\sigma(\lambda^2+\textrm{i}z\lambda^3)t},\nonumber \\
B(t)=\frac{\Omega_{\lambda}-\textrm{i}\Omega_{-\lambda}}{2\sigma(\lambda^2-\textrm{i}z\lambda^3)}\left(1-e^{-\sigma(\lambda^2-\textrm{i}z\lambda^3)t}\right)+B(0)e^{-\sigma(\lambda^2-\textrm{i}z\lambda^3)t},\nonumber 
\eea
while for $\lambda=0$ has solution 
\bea
c_{harm}(t)=c_{harm}(0)+\Omega_{harm}t.
\label{due}
\eea
Note that since $\bm\gamma$ and $\bf \hat{D}$ anticommute, a projection into the right eigenvectors will lead to a change $z\to-z$ in the expressions for  $A(t)$ and $B(t)$ but will leave Eq. (\ref{uno}) and Eq. (\ref{due}) unchanged otherwise.
Therefore the harmonic mode(s) of the signal oscillate(s) freely as dictated by their intrinsic frequencies, while the non-harmonic modes converge to a stable focus, i.e.  {they} converge to a steady value  {and would show} a spiral-in phase portrait.
We therefore find that in the linearized dynamic, LDS  {also} displays emergent (damped) oscillations.

\section{Projected dynamics on fully connected and on random uncorrelated networks}
\label{AppendixB}
In this Appendix  we derive closed form equations for the canonical variables  $\bm\alpha$ and $\bm \beta$ of LDS on fully connected networks. Moreover we derive the equations describing the dynamics of the canonical variables $\bm\alpha$ and $\bm\beta$ on uncorrelated sparse networks  {in the annealed approximation}.
These equations will then be used in Appendix $\ref{AppendixC}$ to derive the phase diagram of LDS for fully connected and sparse networks,  {in the latter case using} the annealed approximation.
\subsection{Dynamics on fully connected networks}
For a fully connected network of $N$ nodes, each node $i$ has degree  $k_i=N-1$ and the adjacency matrix elements are given by $a_{ij} = 1$ for $i\neq j$ and 0 otherwise. For this particular topology, the LDS dynamics (Eq. (\ref{eq1.})) becomes
\bea
\Dot{\theta}_i &= \omega_i + \frac{\hat{c}}{N}\sigma \sum_{j=1}^N\sin{\left(\alpha_j-\alpha_i\right)},\\
\Dot{\psi}_i &=\hat{\omega}_i -\frac{\hat{c}}{2N}\sigma\sum_{j = 1}^N\sin{\beta_j} +\frac{\hat{c}\sigma}{2}\sin{\beta_i},
\label{eq: nodes FC dynamics}
\eea
where $\hat{c} = N/(N-1)$ and where $\alpha_i$ and $\beta_i$ can be expressed as
\bea
\alpha_i& = &\frac{1}{2}\left(\theta_i + z\psi_i\right),\nonumber \\
    \beta_i &= &\frac{z\hat{c}}{2}\left(\theta_i - \Theta\right) - \psi_i,
\eea
with $\Theta = \sum_{i=1}^N \theta_i/N$.
In terms of the two complex order parameters $X_\alpha$ and $X_\beta$,  given by Eq. (\ref{Xab}) the LDS dynamics on a fully connected network can be rewritten as
\begin{align}
    \Dot{\theta}_i &= \omega_i + \hat{c}\sigma\text{Im}\left(X_\alpha e^{-
   \textrm{i}\alpha_i}\right),\\
    \Dot{\psi}_i &=\hat{\omega}_i -\frac{\hat{c}\sigma}{2}\text{Im}\left(X_\beta + e^{-\textrm{i}\beta_i}\right).
    \label{eq:closed}
\end{align}
In order to  {place} ourselves in the rotational frame of the intrinsic frequencies,  we  consider the transformation 
\bea
\alpha \to \alpha - \frac{1}{2}\hat{\Omega}t,
\eea with $\hat{\Omega} = \frac{d\hat{\Theta}}{dt}$.
In  {this} rotating frame,  the closed equations for the phases  $\alpha_i$ and $\beta_i$ are given by 
\bea
    \begin{pmatrix} 
    \Dot{\alpha}_i \\ 
    \Dot{\beta}_i
    \end{pmatrix} = \bm\kappa_i+\frac{\hat{c}\sigma}{2}\text{Im}\left[\mathbf{X}\begin{pmatrix} 
    e^{-\textrm{i}{\alpha}_i} \\ 
    e^{-\textrm{i}{\beta}_i}
    \end{pmatrix}\right],
    \label{eq:closed_ab_FC}
\eea
with 
\bea
\bm\kappa_i&=&\begin{pmatrix} 
    \frac{1}{2}\omega_i + \frac{z}{2}\hat{\omega}_i - \frac{1}{2}\hat{\Omega} - \frac{\hat{c}z}{4}\sigma \text{Im}\left(X_\beta\right)\\
    \frac{z\hat{c}}{2}\omega_i - \hat{\omega}_i - \frac{z\hat{c}}{2}\hat{\Omega} + \frac{\hat{c}}{2}\sigma \text{Im}\left(X_\beta\right) 
    \end{pmatrix}, \nonumber \\
    \mathbf{{X}}&=& \begin{pmatrix} 
    X_\alpha & -\frac{z}{2} \\ 
    \hat{c}z{X}_\alpha & 1
    \end{pmatrix}.
\eea
These equations fully account for the dynamics  of the topological spinor. Hence these equations will be the starting point to predict the phase diagram of LDS on fully connected networks in Appendix $\ref{AppendixC}$.

\subsection{Annealed approximation and dynamics on random networks}
\label{Annealed approximation and dynamics on random networks}

In this section we derive the LDS closed form equations for the canonical variables $\bm\alpha$ and $\bm\beta$ defined on sparse uncorrelated networks, { using} the annealed approximation~\cite{bianconi2002mean,restrepo2005onset}.
Our starting point are the dynamical Eqs. (\ref{eq1.}) where assuming that the LDS dynamics is defined on uncorrelated random networks, we adopt the mean-field annealed approximation \cite{bianconi2002mean,restrepo2005onset}
\bea
a_{ij}\to\frac{k_ik_j}{\avg{k}N},
\eea
i.e. we substitute the generic  adjacency matrix element $a_{ij}$ with the probability of  {occurence of} the link $(i,j)$ in the uncorrelated network ensemble. 
 {We obtain in this way} the dynamical equations
\bea
\dot{\theta}_i&=&\omega_i +\sigma \sum_{j=1}^N 
\frac{k_{j}}{\avg{k}N}\sin \left(\alpha_j-\alpha_i\right),\nonumber \\
\dot{\psi}_{i}&=&\hat{\omega}_{i}-\sigma \frac{1}{2}
\sum_{j=1}^N \frac{k_{j}}{\avg{k}N}\left[\sin \left(\beta_j\right)-\sin \left(\beta_i\right)\right].
\label{eq:annealed}
\eea
Defining the weighted order parameters $\hat{X}_{\alpha}$ and $\hat{X}_{\beta}$ as
\bea
\hat{X}_{\alpha}&=&\hat{R}_{\alpha}e^{\textrm{i}\hat{\eta}_{\alpha}}= \sum_{i=1}^N \frac{k_{i}}{\avg{k}N}e^{\textrm{i}\alpha_i},\nonumber \\
\hat{X}_{\beta}&=&\hat{R}_{\beta}e^{\textrm{i}\hat{\eta}_{\beta}}\sum_{i=1}^N \frac{k_{i}}{\avg{k}N}e^{\textrm{i}\beta_i},
\eea
the annealed Eqs. (\ref{eq:annealed}) become
\bea
    \begin{pmatrix} 
    \Dot{\alpha}_i \\ 
    \Dot{\beta}_i
    \end{pmatrix} = \bm\kappa_i+\frac{\hat{c}\sigma}{2}\text{Im}\left[\mathbf{\hat{X}}\begin{pmatrix} 
    e^{-\textrm{i}{\alpha}_i} \\ 
    e^{-\textrm{i}{\beta}_i}
    \end{pmatrix}\right],
    \label{eq:annealed2}
\eea
where
\bea
\bm\kappa_i&=&\begin{pmatrix} 
    \frac{1}{2}\omega_i + \frac{z}{2}\hat{\omega}_i - \frac{1}{2}\hat{\Omega} - \frac{z}{4}\sigma \text{Im}\left(\hat{X}_\beta\right)\\
    \frac{z}{2}\omega_i - \hat{\omega}_i - \frac{z}{2}\hat{\Omega} + \frac{1}{2}\sigma \text{Im}\left(\hat{X}_\beta\right) 
    \end{pmatrix}, \nonumber \\
    \mathbf{\hat{X}}&=& \begin{pmatrix} 
    \hat{X}_\alpha & -\frac{z}{2} \\ 
    z\hat{X}_\alpha & 1
    \end{pmatrix}.
    \label{kappai}
\eea
The dynamical Eqs.~(\ref{eq:annealed2}) obtained for sparse uncorrelated networks in  the annealed approximation framework can be recast to the closed form equations obtained for fully connected networks (Eq.~(\ref{eq:closed_ab_FC}))  provided that $X_{\alpha}$, $X_{\beta}$ are identified with $\hat{X}_{\alpha}$, $\hat{X}_{\beta}$ and $\hat{c}$ is  {is taken to be} one.

\section{Theoretical prediction of the phase diagram}
\label{AppendixC}

In this Appendix  {we provide} a theoretical framework  {to predict} the phase diagram of LDS on fully connected networks and on sparse networks treated within the annealed approximation.
In particular,  {we focus} on  {this} latter scenario, since the treatment of  LDS on fully connected networks can be easily deduced from derivations obtained within the annealed approximation provided that $X_{\alpha}$, $X_{\beta}$ are identified with $\hat{X}_{\alpha}$, $\hat{X}_{\beta}$ and $\hat{c}$ is  {taken to be} one.
Our starting point  {is} the closed form  Eqs.~(\ref{eq:annealed2}) for the canonical variables $\bm\alpha$ and $\bm\beta$ for LDS on sparse uncorrelated networks.
These equations depend on the vector $\bm\kappa_i=(\kappa_{i\alpha},\kappa_{i\beta})^{\top}$,  {which is a function} on the intrinsic frequencies of the node $i$ under consideration.
The proposed theoretical approach uses as input the intrinsic frequencies $\bm\Omega$, the emergent frequency $\Omega_E$  {as well as} the degree distribution of the network. It provides the phase diagram in the annealed approximation, i.e. the dependence of the real order parameters $R_{\alpha}$ and $R_{\beta}$ on the coupling constant $\sigma$ for every given value of $z$.

The approach is based on a classification of the nodes depending on the dynamical states of the dynamical variables $\bm\alpha$ and $\bm\beta$. In particular, every node $i$  {has} either both phases $\alpha_i$ and $\beta_i$ drifting, or one phase drifting and the other frozen or both phases frozen. This classification of the phases has been proposed in Ref.\cite{calmon2021topological} to treat the phase diagram of Dirac synchronization on a fully connected network and is here extended to treat   LDS on sparse uncorrelated networks within the annealed approximation.

 {We start with} the dynamical Eqs. (\ref{eq:annealed2}) for the phases  {of} a generic node $i$. 
We proceed by classifying the nodes into four classes.
\paragraph{Nodes with frozen phases-}
These are nodes with their phases frozen in the frame oscillating with the emergent frequency $\Omega_E$. Therefore, a generic node $i$ in this class obeys 
\begin{equation}
    \begin{split}
    \dot{\alpha}_i = \Omega_E
    \quad\text{and}\quad
    \dot{\beta} = 0.
    \end{split}   
\end{equation}
Using Eqs. (\ref{eq:annealed2}) it can be shown that this implies that in this frame the phases $\alpha_i$ and $\beta_i$ satisfy
\begin{equation}
    \begin{aligned}
        &\sin(\alpha_i-\hat\eta_\alpha) = d_{i\alpha} = \frac{\omega_i-\hat{\Omega}-2\Omega_E/(1+\hat{c}z^2/2)}{\hat{c}\sigma \hat{R}_\alpha},\\
        &\sin(\beta_i) = d_{i\beta} = \frac{-2\hat{\omega}_i}{\hat{c}\sigma}+ \text{Im}(X_\beta)+2\hat{c}z\frac{\Omega_E}{\sigma \hat{c}(1+\hat{c}z^2/2)}.
    \end{aligned}
\end{equation}
Since the function $\sin(x)$ is bounded by one and minus one, it follows that  only the nodes that admit $|d_{i\alpha}|<1$ and $d_{i\beta}|<1$ belong to this class. The nodes of this class will contribute to all complex order parameters.

\paragraph{Nodes with $\alpha$-oscillating phases-}
The generic node $i$ belonging to this class has $\alpha_i$ phase that oscillates at the emergent frequency $\Omega_E$ while its $\beta_i$ phase is drifting, i.e.
\begin{equation}
    \begin{split}
    \dot{\alpha}_i = \Omega_E
    \quad\text{and}\quad
    \overline{\sin{\beta}_i} \approx 0,
    \end{split}   
\end{equation}
where the overbar denotes a late time average.
Using Eq.~(\ref{eq:annealed2}) it can be shown that the phases $\alpha$ oscillate at frequency $\Omega_E$ so that $\sin (\alpha-\hat{\eta}_{\alpha})$ freezes to 
\begin{equation}
\sin(\alpha_i-\hat{\eta}_\alpha) \approx \hat{d}_{i\alpha} = \frac{2(\kappa_{i\alpha}-\Omega_E)}{\hat{c}\sigma \hat{R}_\alpha},
\end{equation}
as long as $|\hat{d}_{i\alpha}|<1$.
These nodes contribute to the $X_\alpha$ and $\hat{X}_{\alpha}$ order parameters. As their  $\beta$ phase  drifts, these nodes do not contribute to the order parameters $X_\beta$ and $\hat{X}_{\beta}$.

\paragraph{Nodes with $\beta$-oscillating phases-}
These are the nodes $i$ whose $\beta_i$ phase oscillates but their $\alpha_i$ phase is drifting, consequently these phases obey
\begin{equation}
    \begin{split}
    \overline{\text{Im}(\hat{X}_\alpha e^{-\textrm{i}\alpha_i})} = 0
    \quad\text{and}\quad
    \dot{\beta}_i = 0,
    \end{split}   
\end{equation}
which yield the following condition
\begin{equation}
\sin(\beta_i) \approx \hat{d}_{i\beta} = \frac{2\kappa_{i\beta}}{\hat{c}\sigma},
\end{equation}
as long as $|\hat{d}_{i\beta}|<1$.
These nodes contribute to the $X_\beta$ and $\hat{X}_{\beta}$ order parameters only.
\paragraph{Nodes with drifting phases-}
The final class of nodes include all the nodes that do not satisfy any of the previous conditions. The phases of these nodes obey 
\bea
\overline{\mbox{Im}(\hat{X}_{\alpha}e^{-\textrm{i}\alpha_i}})\simeq 0,\quad
\overline{\sin\beta_i}\simeq 0,
\eea
and therefore these nodes do not contribute to any of the order parameters.

In order to derive the expression for ${R}_{\alpha}$ and ${R}_{\beta}$ we note that the proposed classification of the nodes depends on the weighted order parameters and in particular on $\hat{R}_{\alpha}$ and $\hat{X}_{\alpha}$ {. These can} be calculated self consistently using a variation of the approach proposed in Ref. \cite{calmon2021topological},  {giving}
\bea
\hat{R}_{\alpha}&\simeq&\frac{1}{\avg{k}N}\sum_{i=1}^Nk_i\sqrt{1-d_{i\alpha}^2}H(1-d_{i\alpha}^2)H(1-d_{i\beta}^2)\nonumber \\
&&+ \frac{1}{\avg{k}N} \sum_{i=1}^Nk_i\sqrt{1-\hat{d}_{i\alpha}^2}H(1-\hat{d}_{i\alpha}^2)H(d_{i\beta}^2-1),\nonumber \\
\mbox{Im}\hat{X}_\beta &\simeq &\frac{1}{\avg{k}N}
\sum_{i=1}^N k_i{d}_{i\beta}H(1-d_{i\alpha}^2)H(1-d_{i\beta}^2)\nonumber \\
&&+ \frac{1}{\avg{k}N} \sum_{i=1}^Nk_i{\hat{d}_{i\beta}}H({d}_{i\alpha}^2-1)H(1-\hat{d}_{i\beta}^2),
\label{hcomp2ns}
\eea
where here and in the following $H(x)$ indicates the Heaviside function.
 {This allows us to obtain} the expression for the  order parameters ${R}_{\alpha}$ and ${R}_{\beta}$ within the annealed approximation.  {We find that these are} given by 
\bea
R_{\alpha}&\simeq&\frac{1}{N}\sum_{i=1}^N\sqrt{1-d_{i\alpha}^2}H(1-d_{i\alpha}^2)H(1-d_{i\beta}^2)\nonumber \\
&&+ \frac{1}{N} \sum_{i=1}^N\sqrt{1-\hat{d}_{i\alpha}^2}H(1-\hat{d}_{i\alpha}^2)H(d_{i\beta}^2-1),\nonumber\\
R_{\beta}&=&|X_{\beta}|,\label{comp2ns}
\eea
where $\mbox{Im}{X}_\beta$ and $\mbox{Re}{X}_\beta$ can be expressed as 
\bea
\mbox{Im}{X}_\beta &\simeq &\frac{1}{N}
\sum_{i=1}^N {d_{i\beta}}H(1-d_{i\alpha}^2)H(1-d_{i\beta}^2)\nonumber \\
&&+ \frac{1}{N} \sum_{i=1}^N{\hat{d}_{i\beta}}H({d}_{i\alpha}^2-1)H(1-\hat{d}_{i\beta}^2),\nonumber \\
\label{Icomp2ns}
\mbox{Re}{X}_\beta &\simeq &\frac{1}{N}
\sum_{i=1}^N \sqrt{1-d_{i\beta}^2}H(1-d_{i\alpha}^2)H(1-d_{i\beta}^2)\nonumber \\
&&+ \frac{1}{N} \sum_{i=1}^N\sqrt{1-\hat{d}_{i\beta}^2}H({d}_{i\alpha}^2-1)H(1-\hat{d}_{i\beta}^2).
\eea

\end{document}